\documentclass{aa}  

\usepackage{graphicx}
\usepackage{txfonts}
\usepackage[]{hyperref}
\usepackage{xcolor}
\usepackage{soul}
\usepackage{booktabs}
\usepackage{tabularx}
\usepackage{breqn}

\newcommand{\Rer}{\ensuremath{R_{\oplus}}}
\newcommand{\Mer}{\ensuremath{M_{\oplus}}}
\newcommand{\Msun}{\ensuremath{M_{\odot}}}
\newcommand{\Rpl}{\ensuremath{R_{\rm pl}}}
\newcommand{\Mpl}{\ensuremath{M_{\rm pl}}}
\newcommand{\Rroc}{\ensuremath{R_{\rm roche}}}
\newcommand{\HRroc}{\ensuremath{\hat{R}_{\rm roche}}}
\newcommand{\Teq}{\ensuremath{T_{\rm eq}}}
\newcommand{\Feuv}{\ensuremath{F_{\rm EUV}}}
\newcommand{\Mdot}{\ensuremath{\dot{M}}}

\newcommand{\ergscm}{\ensuremath{\rm erg\,s\,cm^{-2}}}

\begin{document} 

   \title{Grid-based exoplanet atmospheric mass loss predictions through neural network}
   \subtitle{}
   \titlerunning{Grid-based mass loss predictions through neural network}
   \author{
   Amit Reza\inst{1} $^{\href{https://orcid.org/0000-0001-7934-0259}{\includegraphics[scale=0.5]{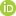}}}$,
   Daria Kubyshkina\inst{1} $^{\href{https://orcid.org/0000-0001-9137-9818}{\includegraphics[scale=0.5]{FIGURES/orcid.jpg}}}$,
   Luca Fossati\inst{1} $^{\href{https://orcid.org/0000-0003-4426-9530}{\includegraphics[scale=0.5]{FIGURES/orcid.jpg}}}$
   Christiane Helling\inst{1} $^{\href{https://orcid.org/0000-0002-8275-1371}{\includegraphics[scale=0.5]{FIGURES/orcid.jpg}}}$
   }
   
   \authorrunning{Reza et al.}
   
   \institute{
    Space Research Institute, Austrian Academy of Sciences, Schmiedlstrasse 6, A-8042 Graz, Austria\\
    \email{amit.reza@oeaw.ac.at}    
   }

   \date{Received XXX; accepted XXX}

 
  \abstract
   {The fast and accurate estimation of planetary mass-loss rates is critical for planet population and evolution modelling.}
   {We use machine learning (ML) for fast interpolation across an existing large grid of hydrodynamic upper atmosphere models, providing mass-loss rates for any planet inside the grid boundaries with superior accuracy compared to previously published interpolation schemes.}
   {We consider an already available grid comprising about 11\,000 hydrodynamic upper atmosphere models for training and generate an additional grid of about 250 models for testing purposes. We develop the ML interpolation scheme (dubbed ``atmospheric Mass Loss INquiry frameworK''; \texttt{MLink}) using a Dense Neural Network, further comparing the results with what was obtained employing classical approaches (e.g. linear interpolation and radial basis function-based regression). Finally, we study the impact of the different interpolation schemes on the evolution of a small sample of carefully selected synthetic planets.}
   {\texttt{MLink} provides high-quality interpolation across the entire parameter space by significantly reducing both the number of points with large interpolation errors and the maximum interpolation error compared to previously available schemes. For most cases, evolutionary tracks computed employing \texttt{MLink} and classical schemes lead to comparable planetary parameters at Gyr-timescales. However, particularly for planets close to the top edge of the radius gap, the difference between the predicted planetary radii at a given age of tracks obtained employing \texttt{MLink} and classical interpolation schemes can exceed the typical observational uncertainties.}
   {Machine learning can be successfully used to estimate atmospheric mass-loss rates from model grids paving the way to explore future larger and more complex grids of models computed accounting for more physical processes.}

   
   \maketitle
\nolinenumbers
\section{Introduction}\label{sec:intro}
The Kepler mission clearly indicated that planets with sizes in between Earth and Neptune (i.e. sub-Neptunes and super-Earths) constitute a significant fraction of the existing exoplanet population \citep[e.g.][]{mullally2015}. Follow-up ground-based radial velocity (RV) observations showed that these planets present an exceptional diversity in terms of planetary bulk density \citep[e.g.][]{otegi2020_MR,mueller2024_MRrelation}. Observations performed with the TESS \citep{ricker2015_tess} and CHEOPS \citep{benz2021_cheops} missions, accompanied by RV measurements aiming at measuring planetary bulk densities with an accuracy typically better than 15\%, have ultimately confirmed the stunning diversity characterising these planets, even when looking within a single planetary system \citep[e.g.][]{gandolfi2019_hd15337,gunther2019_toi270,carleo2020,nielsen2020_toi125,leleu2021_toi178,Lacedelli2021,wilson2022_toi1264,lam2023_toi1260,tuson2023_hd15906,luque2023,bonfanti2023_toi732}.

This large diversity is likely the result of formation and evolution processes \citep[e.g.][]{lee_chiang2017,venturini2020,evelee2022fgap,kubyshkina2022MR}. Among the latter ones, atmospheric loss is believed to play a dominant role \citep[e.g.][]{owen_wu2017,jin2018,modirrousta2020}. Therefore, to be able to accurately model planetary atmospheric evolution, and ultimately interpret observations, we need to adequately estimate atmospheric mass-loss rates. This becomes particularly important when planets go through a phase of hydrodynamic escape, which can be driven by a combination of planetary intrinsic thermal energy and low gravity \citep[``boil-off'' or ``core-powered mass loss'';][]{stoekl2015,owen_wu2017,fossati2017,gupta_schlichting2019}, by the stellar high-energy (X-ray and extreme-UV; together XUV) irradiation \citep[e.g.][]{yelle2004Icar..170..167Y,yelle2008}, and/or tidal forces of the host star \citep[e.g.][]{Koskinen2022ApJ...929...52K,Guo2024NatAs.tmp...89G}. XUV-driven atmospheric escape can be estimated with some accuracy employing the energy-limited formula \citep{watson1981,Erkaev2007}, although it carries severe approximations \citep[e.g.][]{krenn2021}. However, accurately estimating mass-loss rates in the boil-off regime, which is particularly relevant in the early stages of evolution of sub-Neptunes, and in the transition regime between boil-off and XUV-driven escape requires the use of hydrodynamic modelling \citep[e.g.][]{kubyshkina2018grid}.

Due to their slow computational time, hydrodynamic modelling of planetary upper atmospheres is more suitable for studying single systems in detail. Instead, for modelling atmospheric evolution there is the need to find ways to quickly estimate atmospheric mass-loss rates, taking care not to lose accuracy with respect to what is provided by hydrodynamic modelling. To this end, \citet{kubyshkina2018grid} and \citet{kubyshkina_fossati2021} generated a large grid of hydrodynamic planetary upper atmosphere models and developed an interpolation routine capable of estimating mass-loss rates for any planet lying within the grid boundaries. Then, to increase even more the speed at which it is possible to extract mass-loss rates from the grid, \citet{kubyshkina2018app} developed an analytical expression for the atmospheric mass-loss rates based on polynomial fits to the values obtained from the grid. Both interpolation routine and analytical expression have been used by different groups to perform exoplanet atmospheric evolution modelling \citep[e.g.][]{modirrousta2020,kubyshkina2019kepler11,kubyshkina2020mesa,venturini2020,bonfanti2021evol,ketzer2022,affolter2023}.

However, the interpolation routine and the analytical approximation do not enable fast updates of the algorithms following the addition of new grid points. Also, the interpolation routine has a rather low accuracy (i.e. by a factor of a few) in the estimation of mass-loss rates from the grid for planets lying at the transition region between boil-off and XUV-driven escape and for very high equilibrium temperature and low gravity planets. Furthermore, the analytical approximation suffers of similar inaccuracies for massive planets in close orbit to low-mass stars \citep{kubyshkina2018grid,krenn2021}. To overcome these limitations, we present here a new interpolation scheme based on a neural network (NN) machine learning (ML) architecture, which has the further advantage of enabling future quick updates to the algorithm independently of the grid structure and size. 

This paper is organised as follows. Section~\ref{sec:grid} presents a brief overview of the grid of hydrodynamic atmosphere models and the challenges faced by any interpolation algorithm. Section~\ref{sec:test_set} describes the test data set that has been generated with the hydrodynamic code to examine the quality of the interpolation scheme. Sections~\ref{sec:NN}, \ref{sec:NN_grid-app}, and  \ref{sec:testing} describe the employed machine learning algorithm, the tests carried out to validate it, and the characteristics of the new interpolation scheme. Section~\ref{sec:application} presents an example application of the new interpolation scheme in terms of atmospheric evolution modelling, further comparing it to what is provided by the previous algorithms, while Section~\ref{sec:interpol3} describes the usage of the new interpolation routine. Finally, Section~\ref{sec:conclusions} summarises the work presented here and gathers the conclusions.

\section{The grid of hydrodynamic upper atmosphere models}\label{sec:grid}
The original grid of hydrodynamic upper atmosphere models at the basis of this work has been presented by \citet{kubyshkina2018grid} and \citet{kubyshkina_fossati2021}, and each model has been computed using the code described in \citet{kubyshkina2018grid}. Additional hydrodynamic models presented in this work have been computed with the same code. Here, we provide more details on the boundary conditions of the hydrodynamic model, which are important to better understand some of the results described below.

In some specific cases, the choice of boundary conditions can impact the final results of the hydrodynamic modelling, at times in a significant way \citep[e.g.][]{kubyshkina2024cloudy}. Identifying the most adequate set of boundary conditions across the entire grid is not feasible, and thus we decided to fix them homogeneously across the grid and corresponding to what we found being the least impacting on the results, yet enabling the computation of a large grid of models.

The model parameters are fixed at the lower boundary of the simulation domain and are continuous up to the upper boundary (i.e. their radial derivatives equal zero). The lower boundary of the simulation domain is fixed at the planetary photosphere \Rpl\ in the optical band. Further, the atmosphere at \Rpl\ is assumed to be in hydrostatic equilibrium and consisting of molecular hydrogen. The temperature at \Rpl\ is set equal to the equilibrium temperature \Teq\ of the planet (assuming zero albedo) and the pressure to the photospheric values predicted by lower atmosphere structure models assuming solar metallicity \citep{cubillos2017}, which can be approximated as
\begin{equation}\label{eq:PRESSURE0}
    P_{\rm photo} = 0.6\Rpl^{-1.24}\Mpl^{0.62}\left(\frac{\Teq}{300\,{\rm K}}\right)^{-0.47}\,,
\end{equation}
where \Mpl\ is the planetary mass. Thorough testing has shown that changing $P_{\rm photo}$ in the $\sim100$\,mbar -- 1\,$\mu$bar range can lead to differences in the mass-loss rates (\Mdot) within a factor of two for typical sub-Neptune-like planets and up to an order of magnitude for the most extreme hot, low-mass, inflated planets \citep[see also][]{kubyshkina2024cloudy}.

The default position of the upper boundary was taken at the Roche lobe (\Rroc) of the planet along the planet-star line
\begin{equation}\label{eq:Rroche}
    \Rroc = a\left(\frac{\Mpl}{3M_*}\right)^{1/3}\,,
\end{equation}
though, in some cases the position of the upper boundary had to be adjusted. For some close-in planets, the Roche lobe can be as short as a few \Rpl, in which case, the predicted atmospheric outflow can remain subsonic throughout the simulation domain. If this was the case, the upper boundary was extended until it included the sonic point (typically extending it to $\sim1.2-1.5$\,\Rroc\ was sufficient). For long-period planets, the Roche lobe can extend beyond hundreds of \Rpl, while the outflow is fully formed already close to the photosphere. To avoid the unnecessary increase of the radial grid size (and consequently of the computation time), in these cases the simulation domain was limited to 45\,\Rpl\ when \Rroc\ exceeded this value. For these planets, the code automatically checks that the outflow becomes supersonic within the simulation domain and tests performed for various planets indicate that the choice of the upper boundary position has a minor effect on the predictions of the model.
\subsection{Grid structure}\label{sec::grid_structure}
The latest published version of the grid \citep{kubyshkina_fossati2021} includes 10235 model planets, and the range of planetary and stellar parameters is summarised in Table\,\ref{tab:grid_ranges}. For this work, we extended the grid by about 10\%, for a total of 11442 synthetic planets, with the additional runs focusing on regions of the parameter space that required better coverage (see Section~\ref{sec:interpolation_challenges}). Therefore, we describe here the grid structure. 

We considered stars with masses ($M_*$) between 0.4 and 1.3\,\Msun\ and, for each of them, we adopted the X-ray luminosity ranges expected within their main sequence lifetimes \citep[see e.g.][]{jackson2012,Shkolnik2014,matt2015}. The relation between stellar X-ray and EUV luminosities was set using the empirical fit given by \citet{SF2011}
\begin{equation}\label{eq:X-ray_to_EUV}
    \log_{10}(L_{\rm EUV}) = 0.86\log_{10}(L_{\rm X}) + 4.8\,,
\end{equation}
without accounting for the rather large uncertainties of this fit. For each grid point (i.e. synthetic planet), XUV radiation enters the simulation as a flux scaled to the semi-major axis $a$. The range of orbital separations included in the grid is based on the considered range of planetary equilibrium temperatures ($T_{\rm eq}$), which is between 300\,K (roughly the inner edge of the conservative habitable zone) and 2000\,K (to allow for the presence of molecular hydrogen at the lower boundary of the simulation domain), and $M_*$, thus stellar radius ($R_*$) and effective temperature ($T_{\rm eff}$). The $R_*$ and $T_{\rm eff}$ values are derived considering the range of radii and effective temperatures covered by a star of each considered mass along the main sequence on the basis of stellar evolutionary tracks \citep{choi2016}. Therefore, for each star of a given mass and for each specific $T_{\rm eq}$ value (set independently), we set $a$ at the centre of the interval within which the specific $T_{\rm eq}$ is found along the main sequence evolution of the host star.
\begin{table*}
    \centering
    \caption{Parameter ranges of the grid and number of models ($N_{\rm pl}$) for each considered stellar mass.}
    \begin{tabular}{cccccccc}
    \hline
    \hline
      $M_{*}$ & $L_{\rm X}$ & $L_{\rm EUV}$ & $a$ & \Teq\ & \Mpl\ & \Rpl\ & $N_{\rm pl}$\\
            $[\Msun]$ & $\rm{[\log_{10}(erg/s)]}$ & $\rm{\log_{10}erg/s]}$ & $\rm{[AU]}$ & $\rm{[K]}$ & $[\Mer]$ & $[\Rer]$ & \\
            \hline
         0.4 & 23.7--28.25 & 25.2--29.1  & 0.0037--0.0926 & 300--1500 & 1--108.6 & 1--10 & 1149 \\
         0.6 & 23.7--28.9  & 25.2--29.7  & 0.0054--0.2421 & 300--2000 & 1--108.6 & 1--10 & 1767 \\
         0.8 & 24.4--30.8  & 25.8--31.3  & 0.0109--0.4847 & 300--2000 & 1--108.6 & 1--10 & 1905 \\
         1.0 & 24.9--30.1  & 26.2--30.65 & 0.0227--1.0094 & 300--2000 & 1--108.6 & 1--10 & 3138 \\
         1.3 & 24.9--30.6  & 26.2--31.1  & 0.0342--1.5200 & 300--2000 & 1--108.6 & 1--10 & 2276 \\
    \hline
    \end{tabular}
    \label{tab:grid_ranges}
\end{table*}

\begin{figure}
    \centering
    \includegraphics[width=1\linewidth]{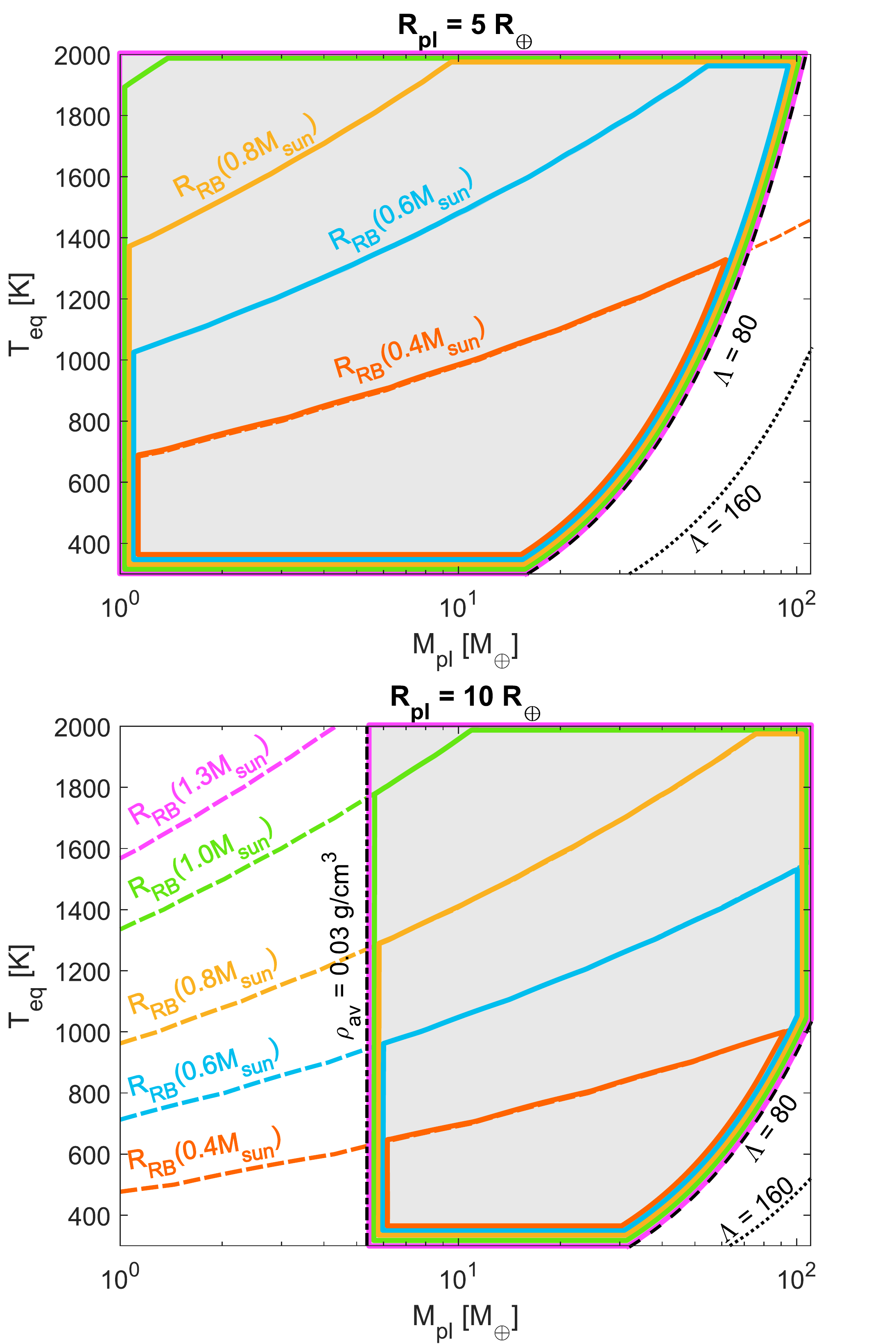}
    \caption{Illustration of the criteria setting the additional boundaries on the grid (see Section~\ref{sec::grid_structure}) in the \Mpl\--$T_{\rm eq}$ plane for $5\,\Rer$ (top) and $10\,\Rer$ (bottom) planets. The orange, blue, yellow, green, and violet dashed lines marked as $R_{\rm RB}(M_*)$ show the upper \Teq(\Mpl) boundaries set by \texttt{criterion II} ($\Rroc\,\geq\,1.2\Rpl$) for planets orbiting 0.4, 0.6, 0.8, 1.0, and 1.3\,\Msun\ stars, respectively. The solid line contours of the respective colours show the grid ranges in the $T_{\rm eq}$--\Mpl\ plane for each of the stellar masses, with all areas sharing the same lower boundary controlled by \texttt{criterion III} (black dashed line for $\Lambda\,=\,80$; black dotted line for $\Lambda\,=\,160$). In the bottom panel, the region on the left of the shaded area (grid ranges) is excluded by the low-density \texttt{criterion I} ($\rho\geq0.03$\,g\,cm$^{-3}$; black dash-dotted line).}
    \label{fig:MTeq_cuts}
\end{figure}

For each pair of $M_*$ and \Teq, there are grid points at 3--6 different values of the stellar X-ray and EUV flux ($F_{\rm X,EUV}$), 20 different values of \Mpl\ in the 1--108.6\,\Mer\ range with a logarithmic spacing, and several values of \Rpl\ in the 1--10\,\Rer\ range. However, not each parameter combination leads to a physically plausible planet and/or leads to a hydrodynamic atmosphere. Therefore, for each combination of $M_*$ and \Teq, we further applied the following three criteria restricting the specific combinations of \Mpl\ and \Rpl\ \citep[see also Figure\,1 of][]{kubyshkina2018grid}. 

\begin{itemize}
\item \texttt{Criterion I}: the grid excludes planets with an average density ($\rho_{\rm av}$) lower than 0.03\,g\,cm$^{-3}$, which roughly corresponds to the lowest density among the planets detected to date \citep[e.g. Kepler-51 system; see][]{masuda2014,hadden2017,Libby-Roberts2020}. This criterion may result being too strict in some cases, e.g. at the very early stages of planet evolution, about the time of protoplanetary disk dispersal, low-mass planets can have even lower densities due to their high post-formation luminosities and consequent radius inflation. However, the occurrence of such low densities depends strongly on the initial entropy of the forming planets, which are poorly constrained, and would last for just a few years due to the planet's fast contraction following the release from the disk.
\item \texttt{Criterion II}: the grid excludes planets where the Roche lobe lies closer than $\sim0.2\,\Rpl$ from the photosphere. For planets with \Rroc\ near the photosphere, the atmospheric outflow is typically independent (or weakly dependent) of photochemistry and occurs in the Roche lobe overflow regime \citep[e.g.][]{Koskinen2022ApJ...929...52K}. In this case, the mass-loss rates predicted by our model depend strongly on the lower boundary conditions.
\item \texttt{Criterion III}: the grid narrows towards high-density planets where the atmosphere is strongly gravitationally bound and can transition into the hydrostatic regime, where our model is not applicable. In the original version of the grid \citep{kubyshkina2018grid}, this criterion had been implemented as a strict boundary at values of the restricted Jeans escape parameter \citep[i.e. the Jeans escape parameter computed at the photosphere;][]{fossati2017}
\begin{equation}\label{eq:Lambda}
    \Lambda = \frac{G\Mpl m_{\rm H}}{k_{\rm b}\Rpl\Teq}
\end{equation}
of 80. In \citet{kubyshkina_fossati2021}, this criterion has been softened to $\Lambda$\,=\,160 to widen the narrow parameter space available for hot planets orbiting low-mass stars (see Figure~\ref{fig:MTeq_cuts}). Namely, if for the original grid we have discarded planets with $\Lambda\,>\,80$ without further consideration, in the later version we model planets with $\Lambda$ values up to 160 if the results for the neighbouring grid points indicate that those planets might not be hydrostatic. This mainly concerns relatively massive ($\Mpl\,\gtrsim\,50$\,\Mer) and compact ($\Rpl\,\lesssim\,5$\,\Rer) planets that can have high $\Lambda$ values even at close orbits (i.e. $\Teq\,\gtrsim\,1000$\,K).

If upon simulation such model planets were found to be hydrostatic (i.e. the exobase level was found to lie below the sonic point), they were not added to the grid. This implies that the actual grid boundaries in terms of $\Lambda$ can differ from $\Lambda\,=\,160$ and lie below it.
\end{itemize}

The three criteria described above manifest differently in different regions of the parameter space. Therefore, the actual ranges of planetary masses/radii present in the grid for different stellar masses and temperatures, and thus orbital separations, can be significantly different. We illustrate this in Figure~\ref{fig:MTeq_cuts} by showing the ranges in the \Mpl--$T_{\rm eq}$ plane for two different planetary radii ($5\,\Rer$ and $10\,\Rer$) and different stellar masses. Within the grid, \texttt{criterion I} is relevant for planets with $\Rpl\,\geq\,5.7\,\Rer$. Therefore, in Figure~\ref{fig:MTeq_cuts} it appears only in the bottom panel ($\Rpl\,=\,10\,\Rer$) and cuts out the planets with masses lower than $\sim5.5$\,\Mer\ (black dash-dotted line). Instead, \texttt{criterion III} is more relevant for compact planets. For example, at $T_{\rm eq}\,=\,300$\,K the condition $\Lambda\,\leq\,80$ excludes planets with masses higher than $\sim$31.7\,\Mer\ for planets of 10\,\Rer, masses higher than $\sim$16.0\,\Mer\ for planets of 5\,\Rer, and masses higher than 3.2\,\Mer\ for planets of 1\,\Rer. Furthermore, \texttt{criterion III} depends on $T_{\rm eq}$, becoming most relevant for cool planets (e.g. for 10\,\Rer\ planets, this criterion is only relevant for planets cooler than $\sim1000$\,K). Finally, \texttt{criterion II}, on top of \Mpl, \Rpl, and \Teq, depends strongly on stellar mass, as planets with the same $T_{\rm eq}$ can orbit stars of different masses at different orbital separations. This criterion is most relevant for hot planets orbiting low-mass stars. In particular, inflated planets orbiting around a 0.4\,\Msun\ star are only available in the grid for temperatures lower than 1360\,K (for 5\,\Rer) or 1000\,K (for 10\,\Rer). Therefore, the three criteria put more stringent constraints on the grid ranges compared to the full ranges listed in Table~\ref{tab:grid_ranges} and lead to significant complications that have to be taken into account when interpolating within the grid of models.
\subsection{Challenges of the grid interpolation}\label{sec:interpolation_challenges}
\subsubsection{The shape of the grid boundaries in six dimensions}\label{sec:interpolation_challenges_boundaries}
The inequality of the mass and radius ranges available in the grid for specific stellar masses and equilibrium temperatures, along with the limited number of grid points (which is dictated by the high computational costs of the hydrodynamic models), poses significant challenges to any grid interpolation. To outline the most problematic regions of the parameter space, panel (a) of Figure\,\ref{fig:LHY-PROCESSES} shows the atmospheric mass-loss rates predicted by the hydrodynamic model for planets with radii of 5\,\Rer\ and masses within the grid ranges orbiting an 0.6\,\Msun\ star at orbital separations corresponding to $T_{\rm eq}$ of 300, 700, 1100, 1500, and 2000\,K as a function of \Mpl. As described above, the minimum and the maximum planetary masses included in the grid are different for each \Teq\ value (i.e. orbit), due to \texttt{criterion II} (minimum \Rroc) and \texttt{criterion III} (maximum $\Lambda$), respectively.
As the quality of the interpolation depends mostly on the density of grid points close to the point of interest, the shortage of points close to the grid boundaries deteriorates the quality of the prediction.
\begin{figure}[ht!]
\centering
\includegraphics[width=0.94\linewidth]{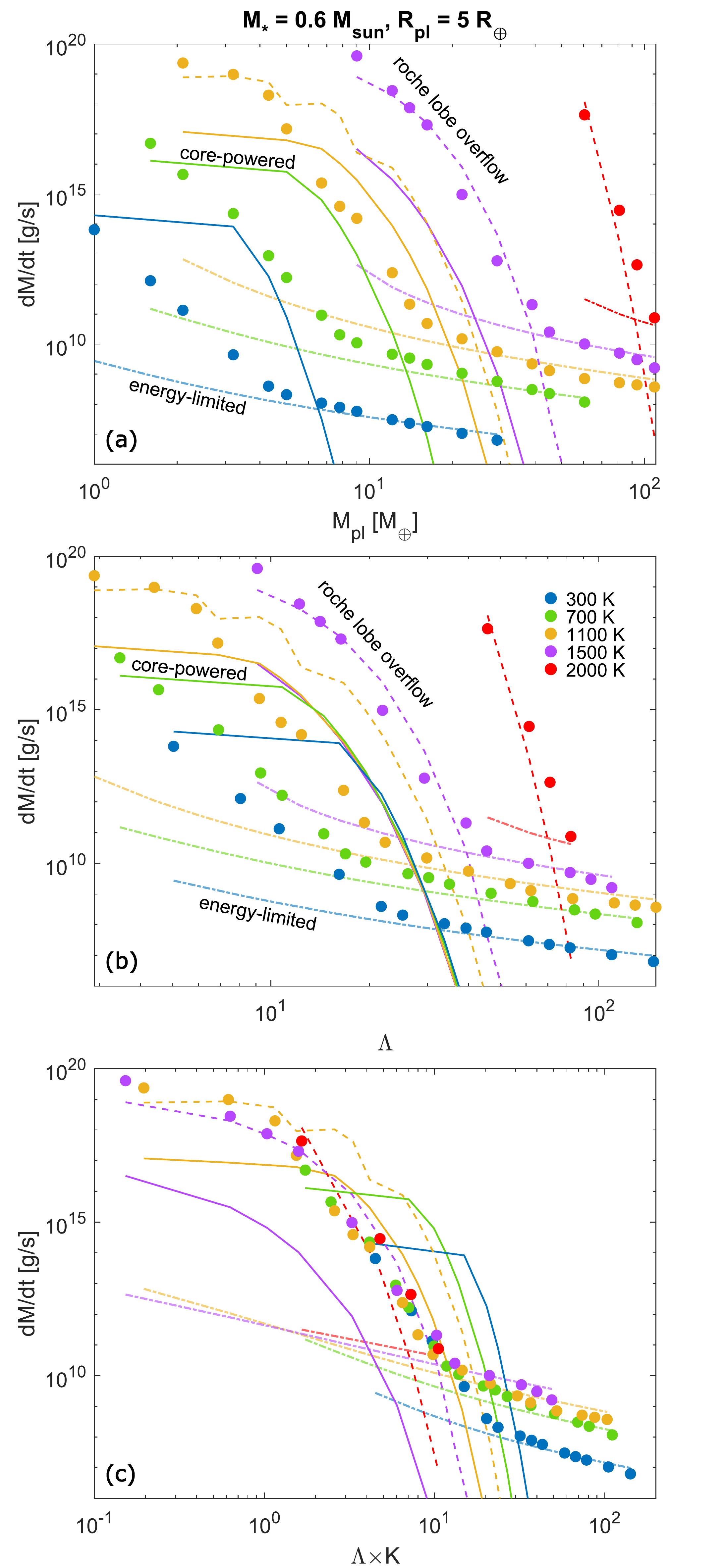}
\caption{Atmospheric mass-loss rates predicted by the hydrodynamic model for planets with $\Rpl\,=\,5\,\Rer$ and \Mpl\ between 1\,\Mer\ and 108.6\,\Mer\ orbiting a $0.6\,\Msun$ star at an orbital separation corresponding to \Teq\ values of 300\,K (blue circles), 700\,K (green circles), 1100\,K (deep yellow circles), 1500\,K (violet circles), and 2000\,K (red circles) as a function of \Mpl\ (top), $\Lambda$ (middle), and $\Lambda\times K$ (bottom). Lines of different styles indicate the mass-loss rates predicted by the energy-limited approximation $\dot{M}_{\rm EL}$ (dash-dotted), core-powered mass loss $\dot{M}_{\rm CP}$ (solid), and Roche lobe overflow $\dot{M}_{\rm RL}$ (dashed), with the colour of the lines following the same scheme as that of the circles. For $\dot{M}_{\rm EL}$, we considered the same EUV fluxes as for the hydrodynamic models shown in the plots, namely 3.0\,\ergscm\ for the orbital separation corresponding to \Teq\ of 300\,K, 89.1\,\ergscm\ for 700\,K, 543.3\,\ergscm\ for 1100\,K, 1878.6\,\ergscm\ for 1500\,K, and 5937.3\,\ergscm\ for 2000\,K.}
\label{fig:LHY-PROCESSES_CLEAN}
\end{figure}

\subsubsection{The switch between different escape regimes occurs across a wide parameter scale}
\label{sec:interpolation_challenges_turnpoint}
Atmospheric escape in the different regions of the parameter space is controlled by various mechanisms and combinations thereof. For hydrodynamic mass loss, these mechanisms are defined by the dominant energy source. For high-gravity planets, the escape is typically driven by photoionisation heating (XUV-driven escape). For hot, low-gravity planets, it can be powered by the planet's internal energy (core-powered mass loss or boil-off). In the extreme cases of planets on very short orbits (i.e. high equilibrium temperatures), the escape is driven by the stellar tidal force (Roche lobe overflow). The transition between the different regimes is illustrated in Figure~\ref{fig:LHY-PROCESSES_CLEAN} and the location at which these regimes transition from one to the other (hereafter called ``regime transition point'') can be identified by where at fixed \Teq\ the mass-loss rate increases significantly with decreasing \Mpl, $\Lambda$, or $\Lambda\times K$. Here, $K$ quantifies the relevance of tidal effects and is defined as \citep{Erkaev2007}
\begin{equation}
\label{eq:K}
K = \frac{(\HRroc-1)^2(2\HRroc+1)}{2\HRroc^3}<1\,,
\end{equation}
where $\HRroc\,=\,\Rroc/\Rpl$. To better show this, we included in Figure~\ref{fig:LHY-PROCESSES_CLEAN} the predictions obtained from three analytical approximations representing the three regimes. For XUV-driven escape, we consider the energy-limited mass loss ($\dot{M}_{\rm EL}$) from \citet[][dash-dotted lines]{Erkaev2007}, for boil-off we consider the core-powered mass loss approximation ($\dot{M}_{\rm CP}$) from \citet[][solid lines]{gupta_schlichting2019}, and for Roche lobe overflow ($\dot{M}_{\rm RL}$) we consider the upper limit given by \citet[][dashed lines]{Koskinen2022ApJ...929...52K}. We describe the exact implementation of each of these approximations in detail in Appendix~\ref{apx:EL-CP-RLO}.

Figure~\ref{fig:LHY-PROCESSES_CLEAN} shows that $\dot{M}_{\rm EL}$ describes well enough the atmospheric mass loss expected for high-gravity planets, but the mass interval in which this match is found depends strongly on \Teq. For example, for the case considered in Figure~\ref{fig:LHY-PROCESSES_CLEAN}, at 300\,K the energy-limited approximation is applicable to planets more massive than $\sim$4\,\Mer, while at 700\,K, 1100\,K, 1500\,K, and 2000\,K, this boundary shifts to $\sim$10\,\Mer, $\sim$20\,\Mer, $\sim$50\,\Mer, and $\sim$110\,\Mer, respectively. Furthermore, for each temperature interval, this boundary shifts towards lower masses for higher EUV fluxes and vice versa (see details in Appendix\,\ref{apx:EL-CP-RLO}).

For lower-mass planets, atmospheric escape is driven mainly by a combination of the planet's low gravity and high internal thermal energy. The shape of the line in Figure~\ref{fig:LHY-PROCESSES_CLEAN} indicating core-powered mass loss comprises a steep component towards higher masses, driven by the assumption that the whole thermal energy of the planet goes into driving the escape, and a flat component, due to the limit on the mass-loss rates set by the thermal velocities of the atmospheric particles at the Bondi radius. For planets with $\Teq\leq1100$\,K, the transition from energy limited to core-powered mass loss occurs near the point where $\dot{M}_{\rm EL}\simeq\dot{M}_{\rm CP}$. Figure~\ref{fig:LHY-PROCESSES_CLEAN} shows that at low \Mpl\ values the core-powered approximation provides just an upper limit to the atmospheric mass-loss rates, while at higher \Mpl\ values and particularly at \Teq\ values higher than about 1000\,K (i.e. short orbital separations) stellar gravity starts driving the escape through Roche-lobe overflow \citep[which occurs when the atmospheric pressure at the Roche lobe is as high as 10--100\,nbar according to][]{Koskinen2022ApJ...929...52K}.
Based on the grid simulations, we find that the tidal forces driving Roche lobe overflow become non-negligible if $\HRroc\lesssim5$ (i.e. $K\lesssim0.7$) and one of the dominant drivers if $\HRroc\lesssim2\pm0.5$ (i.e. $K\lesssim0.3$). The above estimate, however, should not be taken as an exact criterium to classify the escape, as it depends strongly on the system parameters and the assumptions of the hydrodynamic model.

To summarise, there are two regime transition points, one from XUV-driven escape to core-powered mass loss (for planets with longer orbital separations) and one from XUV-driven escape to Roche lobe overflow (for planets with shorter orbital separations). The position of the regime transition point in the \Mpl\ dimension depends on the other system parameters, and particularly on $M_*$ and \Teq. Therefore, if the mass of the planet subject of the interpolation is close to one of the regime transition points, the interpolation can be subject to large inaccuracy, because for such planets the mass-loss rates can change significantly in response to even small variations in one of the planetary parameters, in particular for planets hotter than $\sim1000$\,K.

This problem can be eased by moving from the \Mdot--\Mpl\ space to the \Mdot--$\Lambda$ space. Therefore, instead of interpolating in the 5-dimensional space \{\Mpl, \Rpl, \Teq, \Feuv, $M_*$\}, one can interpolate in the 5-dimensional space \{$\Lambda$, \Rpl, \Teq, \Feuv, $M_*$\}. Combining planetary mass, radius, and temperature (which, in our formulation, define the core-powered mass loss from the planet; see Appendix\,\ref{apx:EL-CP-RLO} for details) within one parameter allows one to significantly decrease the difference between the regime transition points from XUV-driven to core-powered mass loss at different temperatures, as shown in panel (b) of Figure~\ref{fig:LHY-PROCESSES_CLEAN}. However, this does not solve the problem for the hottest planets orbiting around low-mass stars ($\Teq\,\geq\,1500$\,K for 0.6\,\Msun\ and $\Teq\,\geq\,700$\,K for 0.4\,\Msun), as in this case the transition goes from XUV-driven escape to Roche lobe overflow.

To overcome this last obstacle, one can employ $\Lambda\times K$, instead of $\Lambda$ \citep{Guo2024NatAs.tmp...89G}. This approach enables one to almost completely remove the difference between the two regime transition points, as shown by panel (c) of Figure~\ref{fig:LHY-PROCESSES_CLEAN}. In the \Mdot\ vs $\Lambda\times K$ space the regime transition points are located in a very narrow range and their specific position depends mainly on \Feuv\ (they move towards lower $\Lambda\times K$ values for higher XUV fluxes and vice versa). This property can be used to facilitate the interpolation within the grid.
\section{Test dataset}
\label{sec:test_set}
For the development of the previous interpolation algorithms \citep{kubyshkina2018grid,kubyshkina2018app}, the accuracy of the interpolation has been tested by (a) extracting about 10\% of grid points from the whole sample to be then compared with the values given by the interpolation and (b) by comparing the results of the interpolation with the mass-loss rates obtained by directly applying the hydrodynamic model for a dozen real exoplanets. Although both methods have demonstrated to lead to good accuracy of the interpolation \citep{kubyshkina2018grid}, they have drawbacks. While approach (a) provides a sufficiently large test dataset, it implies that all the equilibrium temperatures and stellar host masses of the test planets are those already present in the grid, which essentially reduces the interpolation in the 5-dimensional space to 3-dimensions, namely {\Mpl, \Rpl, \Feuv}. Approach (b), in turn, implies interpolating in the 5-dimensional space, as intended, but lacks in size. Furthermore, the majority of the exoplanets observed to date are rather old, and thus fall into the part of parameter space dominated by XUV-driven mass loss \citep[e.g.][]{kubyshkina2022AN....34310077K}. Therefore, both approaches can turn out to be blind to part of the problematic parameter ranges outlined in Section~\ref{sec:interpolation_challenges}. In this section we describe in detail the sample of simulated planets that is used to test the accuracy of the new interpolation routine described in the next Section.
\subsection{New test planets}
To provide a more realistic estimate of the interpolation errors across the whole parameter space, we built a new test dataset using the same hydrodynamic model used to construct the grid (see Section~\ref{sec:grid}). The test planets were set as follows. For every test planet, we first set the host star mass and equilibrium temperature at random values between 0.4 and 1.3\,\Msun\ and between 300 and 2000\,K, respectively. From these two values, we define the orbital separations ($a\,=\,a(M_*,\Teq)$) in the same way as for the grid planets. We then set the XUV luminosity at a random value within the limits listed in Table~\ref{tab:grid_ranges} for different stellar masses and scale them to the fluxes at the planetary orbits using the corresponding orbital separations. Finally, for the pair of planetary mass and radius, we set \Mpl\ at a random value between 1 and 108.6\,\Mer, using then $M_*$, \Teq, $a$, and \Mpl\ to define the limits for planetary radius employing Criteria I--III described in Section~\ref{sec::grid_structure} and set \Rpl\ at a random value within the derived limits. To set the random values, we used a uniform prior distribution, sampling in the linear space for all parameters except XUV luminosity and \Mpl, which we sampled in logarithmic space. The latter was done so that the parameter distribution of the test sample resembles that of the grid. Furthermore, this approach ensures that planets in the XUV-driven mass loss regime do not dominate the test sample.
\begin{figure}
\centering
\includegraphics[width=0.99\linewidth]{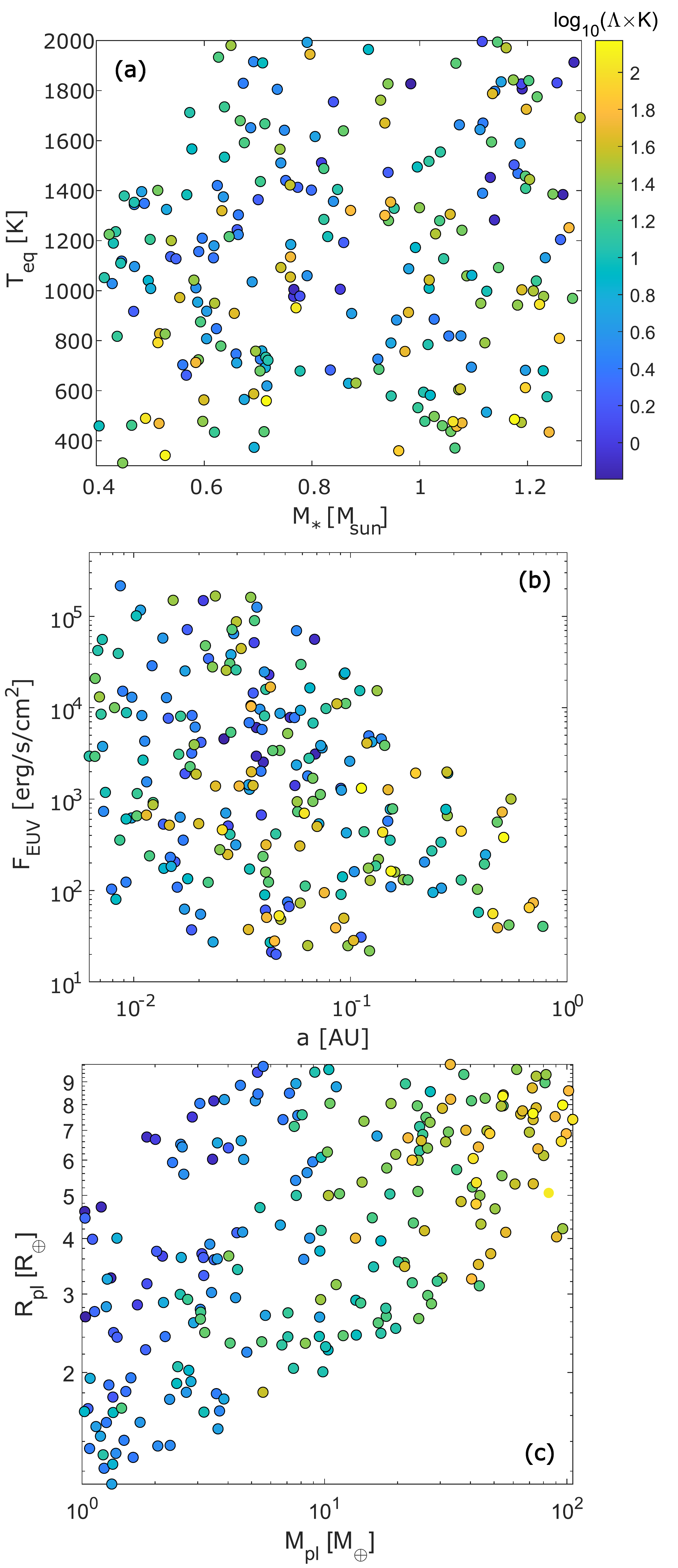}
\caption{Position of the planets comprising the test dataset in the $M_*$--\Teq\ plane (top), in the $a$--\Feuv\ plane (middle), and in the \Mpl--\Rpl\ plane (bottom). In each panel, points are colour-coded according to the value of $\Lambda\times K$ (see the scale in the top panel).}
\label{fig:testset_par}
\end{figure}

After defining the set of test planets, we performed hydrodynamic modelling for each of them. Of the initial 250 models, 16 planets were excluded because the resulting atmospheres did not satisfy the hydrodynamic conditions (e.g. hydrostatic atmosphere), leaving us with a test dataset of 234 planets, which is about 2\% of the entire grid and covers well the grid parameter space. Figure~\ref{fig:testset_par} shows the position of the test planets across the parameter space of the grid.
\subsection{Grid environment of test planets}\label{sec:error_treshold}
To quantify how challenging each of the test planets is in terms interpolation (see Section~\ref{sec:interpolation_challenges}), we consider the following. First, we estimate how close each test planet is to the grid boundaries, to check whether, for example, the mass of the test planet is smaller (larger) than the minimum (maximum) mass among the closest grid points in terms of radius, temperature, and host star mass. Therefore, as any point in the grid can be uniquely defined by five parameters (\Mpl, \Rpl, \Feuv, \Teq, and $M_*$), we look for the grid points closest to each of the test planets. We then fix four of the five parameters to (one of) these values and check if the fifth parameter range available in the grid at this condition covers that of the test planet, and repeat this procedure for different combinations of system parameters. 

To express this numerically, we define the parameter $J_{\rm bord}$ as described in Appendix~\ref{apx:troublemakers}. This parameter reflects the number of cases in which one of the parameters of a test planet lies outside of the boundaries set by the closest grid points. We then normalise $J_{\rm bord}$ so that it can take a maximum value of 1, which corresponds to the situation in which a planet is outside of the grid boundaries considering all parameters. As a guidance, values of $J_{\rm bord}$\,$>$\,0.1 imply that a given test point is not covered by the grid for more than one parameter and at values above 0.15 one can expect the accuracy of the interpolation to drop significantly. For the whole test dataset, $J_{\rm bord}$ lies between 0 and 0.375 and differs from 0 for 166 points out of 234.

\begin{figure}
    \centering
    \includegraphics[width=1.09\linewidth]{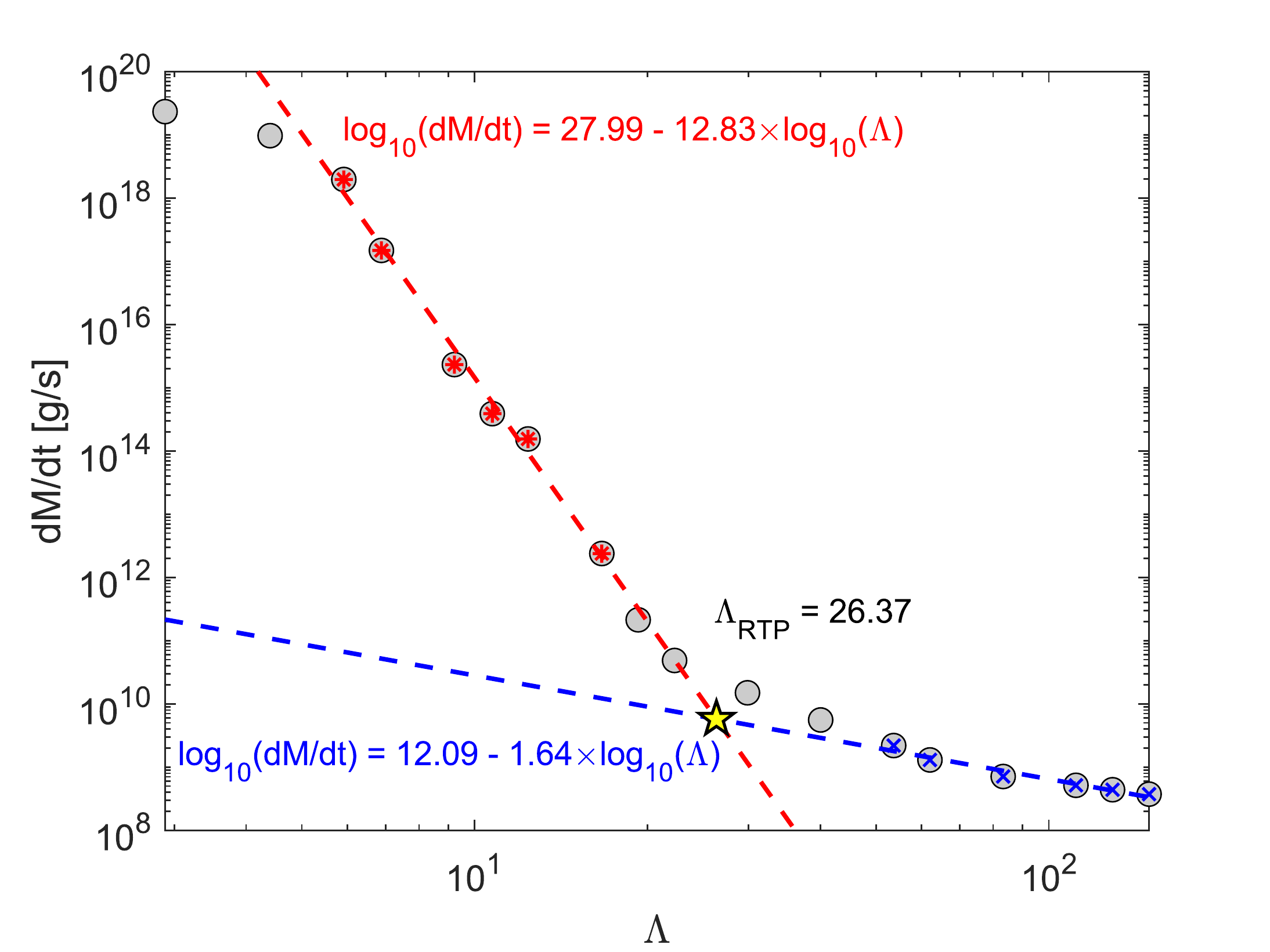}
    \caption{Definition of the location of the regime transition point ($\Lambda_{\rm RTP}$) for $M_*\,=\,0.6$\,\Msun, $\Teq\,=\,1100$\,K, $\Rpl\,=\,5$\,\Rer, and $\Feuv\,=\,543.3$\,\ergscm. Grey circles show the atmospheric mass loss rates against $\Lambda$ given by the hydrodynamic model. The red asterisks denote points identified as planets in the core-powered or Roche lobe overflow regime ($\dot{M}\geq10^{12}$\,g\,s$^{-1}$ and $\Rroc>1.5$; the latter condition excludes points that are too much affected by the lower boundary condition). The blue crosses denote the points in the XUV-driven regime ($\dot{M}\leq10^9$\,g\,s$^{-1}$). The red and the blue dashed lines show the linear fits in the $\log{\dot{M}}$--$\log{\Lambda}$ space for these two groups of points, as specified in the plot. The yellow star gives the position of the regime transition point.}
    \label{fig:fits_RTP}
\end{figure}

In addition to the test planets near the grid boundaries, we identify those lying near the regime transition points. We first locate the position of the regime transition points in terms of $\Lambda$ at fixed $M_*$, \Teq, \Rpl, and \Feuv\ as follows. We identify the points clearly lying in one of the three escape regimes (e.g. within the grid, mass-loss rates above 10$^{12}$\,g\,s$^{-1}$ are reached just through core-powered mass loss or Roche lobe overflow, which we do not distinguish for the determination of the regime transition points). Then, we fit linear functions through the points belonging to each of the regimes in the $\log{\dot{M}}$--$\log{\Lambda}$ space and set the regime transition point at the location where the lines cross (see an example in Figure~\ref{fig:fits_RTP}).

To quantify the distance in $\Lambda$ space between each test point $\Lambda^{\rm i}$ and the regime transition points $\Lambda_{\rm RTP}$, we proceed similarly to the case of $J_{\rm bord}$, as described in Appendix~\ref{apx:troublemakers}. For each $\Lambda^{\rm i}$, we look for the $\Lambda_{\rm RTP}$ values of the nearest neighbours and introduce the parameter $J_{\rm trans}$, which we set equal to one if $|\Lambda_{\rm RTP}-\Lambda^{\rm i}|/\Lambda_{\rm RTP} < 0.2$ (i.e. test point close to a regime transition point) considering any of the nearest neighbours, and to zero otherwise. Then, we look at the points with $J_{\rm trans}\,=\,1$ and re-set them to $J_{\rm trans}\,=\,2$ if the difference between the minimum and maximum values of $\Lambda_{\rm RTP}$ of the considered nearest neighbours exceeds a value of five (a situation similar to the one considered in Section~\ref{sec:interpolation_challenges}). This is done to further highlight points located in regions of the parameter space that are more difficult to accurately interpolate.

\subsection{Performance of the previous interpolation scheme}\label{sec:grid_interpol2021-2023}
\begin{figure*}
\centering
\includegraphics[width=1\linewidth]{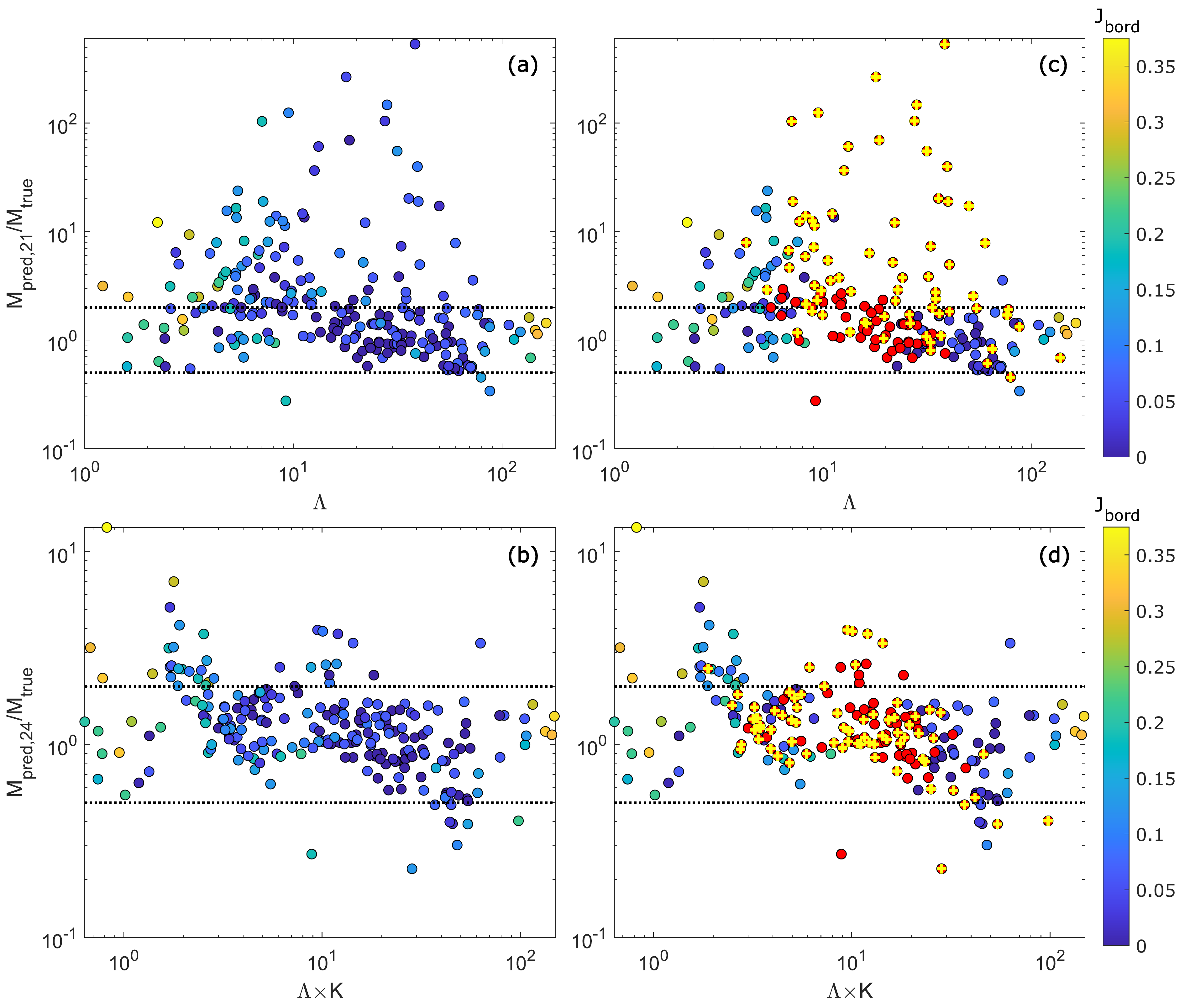}
\caption{Ratio of the mass-loss rates predicted for the test planets by \texttt{interpol2021} (panels a and c) and \texttt{interpol2024} (panels b and d) to the true values given by the hydrodynamic model, against $\Lambda$ and $\Lambda\times K$, respectively. The points are colour-coded according to the $J_{\rm bord}$ value. In panels (c) and (d), the red points have $J_{\rm trans}\,=\,1$, while the red points marked by a yellow plus are those with $J_{\rm trans}\,=\,2$. For reference, the two horizontal black lines in each panel lie at 0.5 and 2.}
\label{fig:testset_predictions_interpol}
\end{figure*}

To roughly estimate the size of the interpolation uncertainties that one can have due to the problems described in Section~\ref{sec:interpolation_challenges}, we calculated the atmospheric mass-loss rates ($\dot{M}_{\rm pred,21}$) for the planets in our test dataset using the old interpolation scheme \citep[\texttt{interpol2021};][]{kubyshkina2018grid,kubyshkina_fossati2021} and then compared them with the estimates obtained using the hydrodynamic model, which we consider here as being the ``true'' values. The results of the comparison are shown in Figure~\ref{fig:testset_predictions_interpol} (panels a and c). We remark that for this test we have disabled the input evaluation procedure included in the distributed version of \texttt{interpol2021}, which would otherwise have discarded all test points with $J_{\rm bord}\,\gtrsim\,0.1$.

To gather an idea of the magnitude of the interpolation uncertainties shown in Figure~\ref{fig:testset_predictions_interpol}, we consider some of the uncertainties intrinsic to the modelling approach and resulting from some of the assumptions. Besides fundamental assumptions (e.g. hydrogen-dominated composition, cloud-free atmosphere, cross-section values), it is difficult to properly quantify the modelling uncertainties, and thus here we focus on two aspects that we believe drive most of the uncertainties, namely the assumptions on the boundary conditions and the treatment of the stellar XUV radiation. Specifically, testing indicates that for the majority of sub-Neptune-like planets in the XUV-driven escape regime on average the assumption taken to set the position of the lower boundary leads to mass-loss rate uncertainties of about a factor of two \citep[see also][]{kubyshkina2018grid,kubyshkina2024cloudy}, though for the most extreme planets in the grid ($\Lambda\times K\,\leq\,1$) it can reach up to a factor of 100. For what concerns the treatment of the stellar XUV irradiation, atmospheric heating depends on the depth at which photons are absorbed, and thus on the shape of the stellar XUV spectrum, which can vary significantly from star to star. However, the grid does not take into account the specific shape of the stellar XUV emission. Furthermore, the model used to compute the grid simplifies the shape of the stellar XUV irradiation representing it in just two wavelength points (i.e. $2\lambda$-approximation; X-ray and EUV). These two simplifications affect the mass loss predictions for most of the intermediate-mass planets ($\Lambda\times K\,\sim\,20-50$). Also in this case, the mass-loss rate uncertainties associated with these two simplifications are expected to be within a factor of two \citep[e.g. ][]{guo_benjaffel2016,Guo2024NatAs.tmp...89G,kubyshkina2022_3d,kubyshkina2024cloudy}. In summary, we consider ratios of $\dot{M}_{\rm pred}/\dot{M}_{\rm true}$ lying between 0.5 and 2 to be accurate predictions.

For our test dataset, in the case of \texttt{interpol2021} about 40\% of the points (95) fall out of this range. Of them, 37 points have $J_{\rm bord}\,>\,0.1$ indicating that they lie near the grid boundaries in their respective parameter regions, and 61 points are close to a regime transition point ($J_{\rm trans}\,>\,0$). Indeed, the majority of the test points with large interpolation errors have $J_{\rm trans}\,=\,2$, and the error is maximum at the intermediate $\Lambda$ values (7--40). Thus, the most problematic planets for the interpolation are hot planets orbiting low-mass stars near a regime transition point.

In an effort to improve the interpolation quality for such planets, we started by revising the previous interpolation scheme \texttt{interpol2021}, in which the interpolation in five dimensions \citep[which did not provide satisfactory results for the tests described in ][]{kubyshkina2018grid} is reduced to a series of one dimensional (1D) interpolations. Furthermore, \texttt{interpol2021} interpolates over the $\Lambda$ values instead of planetary mass and picks one of the pair (\Teq, $a$) as a guiding parameter depending on host star mass and temperature (\Teq\ has a stronger impact on the mass-loss rates for cooler planets around more massive stars and $a$ dominates for hot planets around low-mass stars). 

As the new $\Lambda\times K$ parameter performs better than $\Lambda$ at synchronising the regime transition points in different parameter intervals, we decided to switch from interpolating across $\Lambda$ to interpolating across $\Lambda\times K$. This allowed us to exclude $a$ from the interpolation parameters and switch to \Teq, only. Furthermore, we changed the interpolation on \Teq\ from linear to logarithmic. At last, we narrowed the criteria for picking the neighbouring points used for interpolation. We refer to this version of the interpolation routine as \texttt{interpol2024} and show the testing results analogous to the case of \texttt{interpol2021} ($\dot{M}_{\rm pred,24}/\dot{M}_{\rm true}$) in panels (b) and (d) of Figure~\ref{fig:testset_predictions_interpol}.
    
Considering the test dataset, the use of \texttt{interpol2024} allowed us to reduce the total number of outliers to 45 ($\sim$20\% of the test planets; 26 points correspond to $J_{\rm bord}\,>\,0.1$ and 18 to $J_{\rm trans}\,>\,0$) and the maximum interpolation error by about 43 times, compared to \texttt{interpol2021}. However, there is still some systematic behaviour in the interpolation predictions, which tends to overestimate mass loss at low $\Lambda\times K$ values and to underestimate it at high $\Lambda\times K$ values. This is why we explored other interpolation algorithms, in particular, based on ML.

\section{Exploring the grid interpolation in six dimensions}
\label{sec:NN}
As described in Section~\ref{sec::grid_structure}, in the grid of hydrodynamic models the orbital separation is a function of equilibrium temperature and stellar mass. Therefore, each planet in the grid can be identified by five unique parameters \{$M_*$, \Feuv, \Rpl, \Mpl, \Teq\ or $a$\}. However, the value of the atmospheric mass-loss rate can have a stronger dependence on \Teq\ than $a$ in some regions of parameter space (e.g. for long-period planets), and vice versa in others (e.g. short-period planets). To take advantage of the learning abilities of ML, in the following, we are oblivious to the degeneracy between \Teq\ and $a$ and treat them as independent input parameters.

Considering mass-loss rates as a function of six parameters \{${M}_{*}$, \Feuv, ${a}$, $\Teq$, $\Mpl$, $\Rpl$\} allows one to define the relationship between these parameters and the mass-loss rate as a function $f$ mapping from six-dimensional parameter space to one dimension space ($f:\,\mathbb{R}^{6}\,\rightarrow\,\mathbb{R}$) in a way that $y\,=\,f(\vec{x})$, with $\vec{x} \in \mathbb{R}^{6}$, and $y \in \mathbb{R}$. The unknown function $f$ can be approximated via data-driven modelling and then applied to predict the mass-loss rate $y^{(q)}$ at a new query point $\vec{x}^{(q)}$, such that $y^{(q)} = f(\vec{x}^{(q)})$. Obtaining a numerical function from a set of data points is a high-dimensional (here, six-dimensional) non-linear regression problem that can be solved using several numerical schemes. Here, we mainly focus on two methods: traditional interpolation in the form of Radial Basis Function \citep{buhmann2000radial, powell2001radial} and ML model in the form of NN \citep{hecht1992theory, gurney2018introduction, abiodun2018state, wu2018development}.
\subsection{Neural network-based regression}
\label{sec:NN_method}
We aim to design a NN architecture learning the relationship ($f$) between inputs $\vec{x}\,=\,(M_*,\,\Feuv,\,a,\,\Teq,\,\Mpl,\,\Rpl$) and output $y\,=\,\dot{M}$. We explored several neural network models, including Dense Neural Networks (DNN), Convolutional Neural Networks (CNN), and Recurrent Neural Networks (RNN), to train our data and evaluate prediction performance with the test data points described in Section~\ref{sec:test_set}. The DNN model, consisting of fully connected layers, is well-suited for data with relatively low input and output dimensions. DNN is effective at learning complex mappings between inputs and outputs when the relationships in the data do not require sequential processing.

CNN \citep[CNN;][]{wu2017introduction, albawi2017understanding, gu2018recent} and RNN \citep[RNN;][]{pascanu2013construct, salehinejad2017recent} could offer better solutions with higher-dimensional data (both input and output data are high-dimensional arrays). Convolutional layers in a CNN model extract features from higher-dimensional data, which helps extract hidden features. Since our data do not require extracting hidden features, adding a convolution layer to the input layer is not advantageous. In fact, our tests showed that CNN-based predictions of mass-loss rates from six-dimensional grid points deviated significantly from the actual values, aligning with expectations for CNN performance in cases like ours.

The RNN, including variants such as Long Short-Term Memory \citep[LSTM;][]{hochreiter1997long} networks, are specific types of NN models, where the input of the current layer is connected to the output of the previous one. This allows the model to create a memory through a chain of relations between the input and output of the hidden layers. Thus, RNN models help to better forecast specific kinds of data, such as time series. Although RNNs can predict mass-loss rates with accuracy comparable to DNNs under specific configurations, they come with a higher computational cost due to the increased number of trainable parameters. Furthermore, RNN requires significantly larger training datasets to outperform DNN in scenarios like ours. The mass-loss rate prediction obtained using a simple RNN and LSTM model is shown in Appendix~\ref{apx:lstm}.

Since CNN and RNN have more trainable parameters than DNN, they require significantly larger training datasets to achieve better prediction quality. Furthermore, the simplicity and the ability to generalise well from limited datasets allow DNN to achieve better prediction accuracy than advanced and complex models like CNN or RNN, making the DNN an optimal choice. Given the training dataset size and the nature of the parameter space, the DNN model offers superior prediction accuracy, making it the most suitable choice for our analysis. Hence, our work mainly focused on DNN model-based prediction.

\begin{figure}[ht!]
\centering
\includegraphics[trim={0 0 0 0}, clip, scale = 0.8]{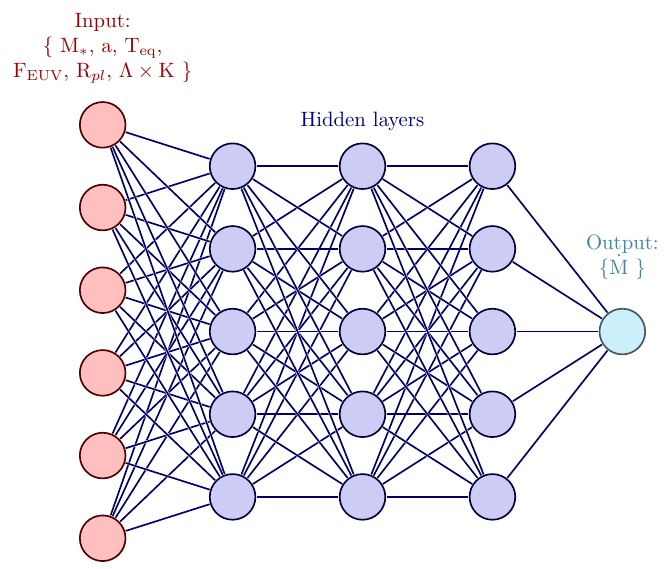}
\hfill
\caption{Schematic representation of a DNN architecture connecting the input parameters and mass-loss rate (output) through a series of hidden layers. In this work, we finally opted for three hidden layers and eight neurons.}
\label{fig:NN-architecture}
\end{figure}

A DNN is a simple NN architecture with dense (fully connected) layers. It is one of the primary and fundamental architectures in deep learning for solving linear and non-linear regression problems. In this type of network, each neuron in a layer is connected to every neuron in the subsequent layer. The network typically consists of an input layer, one or more hidden layers, and an output layer (see Figure~\ref{fig:NN-architecture}). The input layer receives the data, and each hidden layer processes this data by applying a linear followed by a non-linear transformation. The linear transformation of the data is done through weights and biases, while the non-linear transformation is obtained via the activation function \citep{narayan1997generalized, sibi2013analysis, sharma2017activation, ramachandran2017searching}.
The activation function can, for example, be a $\texttt{Sigmoid}$, a Rectified Linear Unit ($\texttt{ReLU}$), an Exponential Linear Unit ($\texttt{ELU}$), a $\texttt{Tanh}$. Transforming the whole training input data through linear--non-linear transformation is called forward propagation. The output layer provides the final predictions, with the number of neurons in this layer corresponding to the number of prediction regression targets (in our case, the single parameter $\dot{M}$). The loss function measures the correctness of the projections. For the regression problem, the loss is usually defined as the mean-squared error (MSE) between predicted and true values. With initial epochs of NN training, the loss is expected to be high. Hence, during training, the network updates weights and biases via back-propagation and gradient descent, which minimises the loss function. Despite its simplicity, a network with dense layers can capture complex patterns in the data, making it a powerful tool for solving non-linear regression problems.

%
\subsection{Radial basis function-based regression} 
\label{Sec:RBF-Regression}
The regression problem can also be solved through traditional interpolation methods such as Chebyshev polynomials and the Radial Basis Function (RBF). Chebyshev polynomials require a fixed grid structure for interpolation, whereas RBF is a mesh-free method. Due to the mesh-free nature of the RBF method, the number of interpolation nodes can be scaled with increasing dimensions of the parameter space. Hence, RBF-based regression is suitable for high-dimensional interpolation with a small/limited number of nodes, which is the case for our grid of hydrodynamic models. The number of required nodes for Chebyshev's polynomial scales exponentially with increasing dimension number of the parameter space, which increases computational costs. Hence, we restricted ourselves to exploring the regression problem with RBF only. Mathematically, an unknown function between a given set of input $\vec{x}$ and output $y$ can be written as a linear combination of N-specific radial basis and unknown weights.
\begin{equation}
f(\vec{x}) = \sum_{i = 1}^{\text{N}}{w_{i} \, \Phi_{i}(\| \vec{x} - \vec{x}_{i} \|) }\,,
\label{Eq:RBF}
\end{equation}
where $\{w_{i}, i = 1, 2, \cdots, \text{N} \}$ are the adjustable weights (or coefficients) and $\{\Phi_{i}\}$ are the radial basis functions, which depend on the distance between the input $\vec{x}$ and a centre $\vec{x}_{i}$. Equation~\ref{Eq:RBF} shows that in an RBF-based regression problem, the given input data $\{ \vec{x}_{i}\}$ is transformed using RBF-basis (kernel), which can be represented by known functions (e.g. linear and Gaussian). Hence, the regression problem boils down to an optimisation in which we need to optimise the corresponding unknown weights of the RBF basis.     
\section{Application to the grid of upper atmosphere models}
\label{sec:NN_grid-app}
The treatment of RBF and NN models for performing regression tasks differs significantly due to their inheritance structure of the solution schemes. NN is a ML method, where training and testing stages are involved, which typically means that the available data (in our case, grid points) is split into training and testing subsets (e.g. 90\% vs 10\% of the points), where only the former one is used to optimise the solution. Meanwhile, in RBF a system of equations is solved to obtain the optimised weights without splitting the data between training and testing. However, to compare the performance of the models, we used the entire grid points in both cases, i.e. to train the NN model or to obtain optimal weights of radial basis, and employed the test dataset described in Section~\ref{sec:test_set} to test the performance of both NN and RBF.  

The results obtained for the old interpolation scheme indicate that the form we use to represent the basic input parameter can affect the results.
Therefore, we explored several parameter combinations to solve the regression problem using NN and RBF. Along with the usual six-dimensional parameter space with \{${M}_{*}$, $\Feuv$, ${a}$, $\Teq$, $\Mpl$, $\Rpl$\}, we tried to solve the regression problem using other combinations, including
\begin{itemize}
\item parameter space reduced to 5 dimensions by excluding \Teq\ or $a$,
\item six-dimensional parameter space, but with \Mpl\ replaced by $\Lambda$ or $\Lambda \times \text{K}$,
\item parameter space extended to 7 dimensions, including both \Mpl\ and $\Lambda$ or $\Lambda \times \text{K}$,
\end{itemize}
and combinations thereof. We found that solving the regression problem with the six-dimensional input parameters \{${M}_{*}$, $\Feuv$, ${a}$, $\Teq$, $\Lambda \times \text{K}$, $\Rpl$\} (namely with $\Lambda \times \text{K}$ replacing \Mpl), performs better in predicting the mass-loss rate for the test data set as compared to other options.

For the training procedure of the DNN model, we used the entire grid, namely 11442 points. The input variables \Teq\ and \Feuv\ were taken on a base-$10$ logarithmic scale. The mass-loss values are taken as base-$10$ logarithmic values and normalised such that $\max{(\log_{10}(\dot{M}))}\,=\,1$ by applying $\texttt{MinMaxScaler}$ scale. 
Therefore, the final prediction of the DNN is between zero and one, and we perform two reverse transformations to obtain the final prediction in the original scale. 

\begin{figure}[!ht]
\centering
\includegraphics[trim={0 0 0 0}, clip, scale = 0.74]{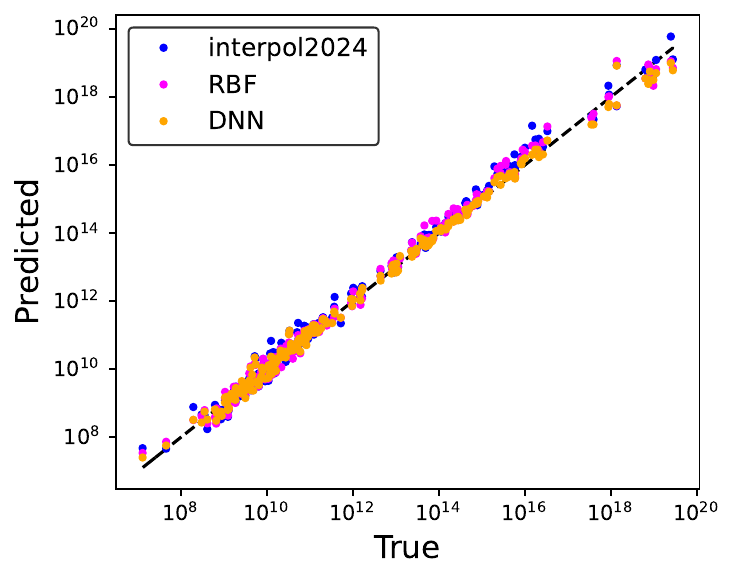}\hfill
\vspace{-0.3cm}
\caption{Mass-loss rate values predicted by the DNN (dark yellow), RBF (magenta), and \texttt{interpol2024} (blue) schemes in comparison to the true values obtained from the hydrodynamic modelling for the test data set. The prediction accuracy of DNN, RBF, and \texttt{interpol2024} are roughly comparable (see Section~\ref{sec:testing}), and outperform those of \texttt{interpol2021}.}
\label{fig:NN_RBF_vs_interp}
\end{figure}

Due to the limited training dataset, investigating the correct choice of an NN architecture is crucial to avoid overfitting and achieve a good generalised functional form. Overfitting occurs when the DNN has too many trainable parameters (such as weights and biases), enabling it to fit the training data closely, but resulting in poor performance on unseen test data \citep{hinton2012improving}. Conversely, a model with fewer parameters can lower the risk of overfitting, but it must retain sufficient capacity to capture meaningful information, ensuring it generalises effectively to new data. The parameters for each dense layer are calculated based on the number of input units, weights, and biases. For example, in this problem, six is the dimension of the input vector, and $\ell$ is the number of neurons. There will be a weight matrix of size $6 \times \ell$ and $\ell$ number of biases. Thus, the parameters between the input layer and the first hidden layer are $7 \ell$ ($= 6 \times \ell + \ell$). Similarly, the number of trainable parameters between the first and second hidden layers is $9 \ell$ ($= 8 \times \ell + \ell$). Hence, the total number of trainable parameters ($N_{\rm TP}$) for a fully connected DNN is
\begin{equation}\label{eq:DNN_num_par}
N_{\rm TP} =  \sum_{i = 1}^{h_{k}} {(m+1) \times n}\,,
\end{equation}
where $m$ and $n$ are the numbers of input and output units for each layer, respectively, and $h_{k}$ is the number of hidden layers. The plus one term is to include the bias terms. Equation~\ref{eq:DNN_num_par} indicates that many hidden layers and a large number of neurons increase the number of total trainable parameters that, in turn, increase the complexity of the problem, and more training data would be needed. For simplicity, we begin with a lighter DNN architecture (e.g. a single hidden layer), gradually increase complexity in terms of the number of layers (up to 10) and neurons (up to 512; see Appendix~\ref{apx:lstm}), and obtain prediction accuracy on the test data set for each architecture. The training of the models has been carried out for a maximum of 5000 epochs, with a learning rate of $10^{-4}$ and batch size of 32. Out of all variations, we found that three layers of DNN architecture with $\texttt{tanh}$ activation function and eight neurons provide the best prediction accuracy on the unseen test data (described in Section~\ref{sec:test_set}), as shown in Figure~\ref{fig:NN_RBF_vs_interp}. 

\section{Validation of the new approach}
\label{sec:testing}
\begin{figure*}[!ht]
\centering
\includegraphics[trim={0 0 0 0}, clip, scale = 0.82]{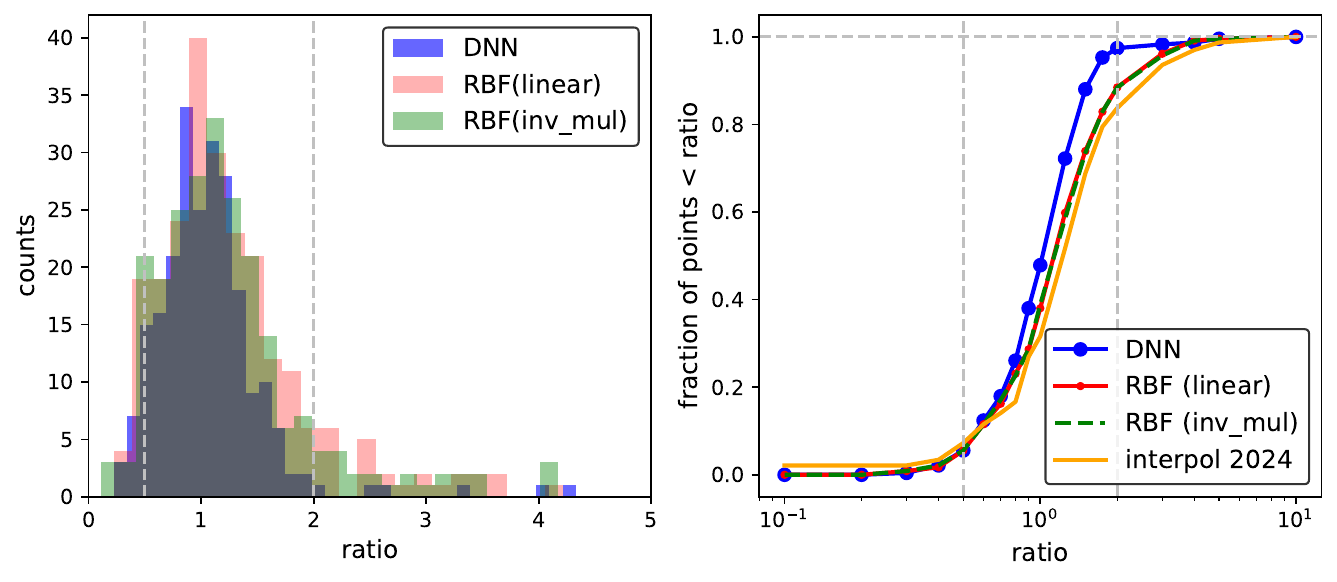} \hfill
\vspace{-0.3cm}
\caption{Distribution of the mass-loss rate ratio between interpolated and true values for the test data set. Left: Ratio between interpolated and true values for the test data set using DNN (blue), RBF with $\textit{linear}$ kernel (red), and RBF with $\textit{inverse_multiquadratic}$ kernel (green). For DNN, $\sim$92\% of the points have ratios lying between 0.5 and 2, while for RBF and \texttt{interpol2024} (not shown for better visualisation) it is $\sim$83\% and $\sim$78\%, respectively. Right: Cumulative distribution of the fraction of points having the mass-loss rate ratio in the 0.1--10 range. DNN interpolation outperforms the other methods. The RBF interpolation with $\textit{linear}$ and $\textit{inverse_multiquadratic}$ ($\text{inv_mul}$) are nearly identical. The choice of shape parameter $\epsilon$ and degree of polynomials are different and have been obtained via a grid-based optimisation process, as described in Appendix~\ref{apx:rbf-kernel}.
}
\label{fig:Err_NN_vs_interp}
\end{figure*}

To test the effectiveness of the trained DNN and RBF models, we used the test dataset of 234 planets described in Section~\ref{sec:test_set}, which is unknown to the trained model (i.e. input grid). We computed the ratio between actual and interpolated mass-loss rates considering DNN and RBF. The left panel of Figure~\ref{fig:Err_NN_vs_interp} shows the histogram of this ratio, which is expected to be around one if the interpolation matches the actual value. For DNN and RBF, the peak of the histogram is nearly at one, showing that for most of the test data set the interpolation is reasonable. For both models, the largest ratio is of order five and is found for a specific point only, while, for most of the points, the ratio lies between 0.5 and 2. 

The right panel of Figure \ref{fig:Err_NN_vs_interp} shows the fraction of points having the mass-loss rate ratio in the 0.1--10 range. The number of points lying in this ratio range is $\sim$92\%, $\sim$83\%, and $\sim$78\% for DNN, RBF, and $\texttt{interpol2024}$, respectively, reflecting the better performance of DNN over the other two methods. Appendix \ref{apx:rbf-kernel} describes in detail the RBF kernel optimisation that has been carried out.

\begin{figure*}[!ht]
\centering
\includegraphics[trim={0 0 0 0}, clip, scale = 0.92]{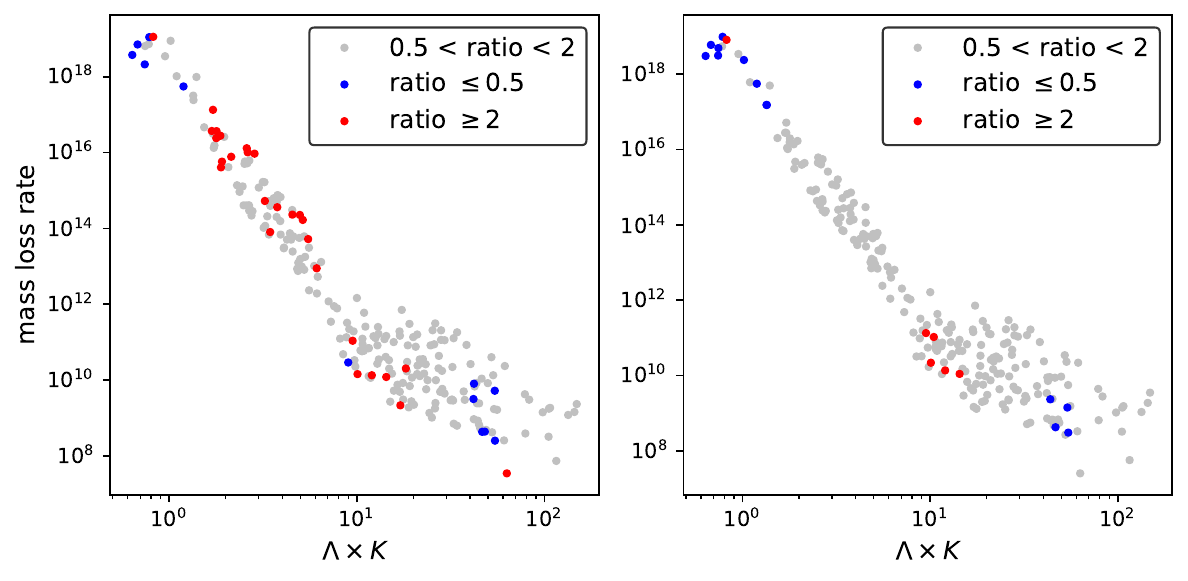} \hfill
\vspace{-0.3cm}
\caption{Mass-loss rate of the test data set as a function of $\Lambda \times \text{K}$ highlighting in red and blue the points for which the ratio between interpolated and true mass-loss rates is smaller than 0.5 or larger than 2, respectively, considering the RBF (left) and DNN (right) interpolation algorithms.}
\label{fig:all_pred_plot}
\end{figure*}

Figure~\ref{fig:all_pred_plot} further describes the quality of the interpolation for the points comprising the test data set as a function of the interpolation algorithm. DNN leads to the smallest number of points in which the interpolated value lies away from the true value by more than a factor of two. As expected, these points are located at the edges of the grid in terms of mass-loss rate and close to the regime transition points. Therefore, future grid extensions should also focus on increasing the number of points in these regions of the parameter space, which will then lead to improve the quality of the DNN interpolation.

\section{MLink: atmospheric Mass Loss INquiry frameworK}
\label{sec:interpol3}
We provide public access to the trained neural network and RBF models, \footnote{\href{https://www.oeaw.ac.at/en/iwf/forschung/forschungsgruppen/exoplaneten-charakterisierung-und-evolution/open-source-software}{https://www.oeaw.ac.at/en/iwf/forschung/forschungsgruppen/\\exoplaneten-charakterisierung-und-evolution/open-source-software}} allowing users to directly predict a mass-loss rate without training the model from scratch. The neural network has been trained using the TensorFlow library and saved in the HDF5 format, which is compatible with TensorFlow's $\texttt{load_model}$ function. Users can load the pre-trained model into their environment to predict outcomes on new datasets with minimal setup. This functionality is included in both command-line and plug-in versions, ensuring users can leverage the predictive capabilities of the model in their preferred workflow. Detailed instructions for loading and using the pre-trained model are provided in the documentation, ensuring a smooth and efficient user experience.

\noindent\textbf{Command-line version:} This version allows users to run the model directly from the command line, making executing the testing processes straightforward. \\

\noindent\textbf{Plug-in version:} This version is tailored for integration within a user-defined modelling framework (e.g. the atmospheric evolution model described in Section~\ref{sec:application}). It includes additional functionality to interface with an external code, allowing the model to adapt dynamically and evolve based on incoming data. This version is designed with modularity in mind, ensuring it can integrate into existing systems.

\section{Impact of improved accuracy on exoplanetary evolution calculations}
\label{sec:application}
Atmospheric evolution modelling requires re-estimating the atmospheric mass-loss rates thousands of times throughout the planetary lifetime, which, due to its fast computation, is the ideal application test-bed for the new interpolation scheme presented above \citep[see e.g.][]{kubyshkina2019kepler11,kubyshkina2020mesa,bonfanti2021evol,affolter2023}. Furthermore, the impact of inaccuracies in the estimation of the mass-loss rates is expected to accumulate in presence of systematic errors. In this section, we use the atmospheric evolution framework developed by \citet{kubyshkina2020mesa,kubyshkina2022MR} to present tests of how different interpolation schemes affect the planetary evolutionary tracks. To complete the picture, we compare the evolutionary tracks obtained using \texttt{interp2024} and \texttt{MLink} to those obtained using the analytical approximations commonly employed to estimate the atmospheric mass-loss rates (energy-limited equation and core-powered mass loss, as described in Appendix~\ref{apx:EL-CP-RLO}). In particular, we employ the evolution model from \citet{kubyshkina2022MR}, which is based on the earlier framework by \citet{chen_rogers2016} and combines the thermal evolution of planets (lower atmosphere structure models) performed with Modules for Experiments in Stellar Astrophysics \citep[MESA][]{paxton2011,paxton2013,paxton2015,paxton2018} code with the atmospheric mass loss at the upper boundary prescribed by \texttt{MLink}, \texttt{interpol2024}, or one of the analytical approximations.

We consider planets in the Neptune-like range of masses (initial masses $M_{\rm pl}^0$ of 5\,\Mer, 9\,\Mer, 17\,\Mer, and 30\,\Mer) evolving in orbits corresponding to the temperature range of 500--1700\,K (where the atmospheric escape is significant, hence, the impact of mass loss is largest) around stars of 0.5, 0.7, and 0.9\,\Msun. The temperature and stellar mass values correspond to the parameter regions where the interpolation errors of our method are expected to be largest. For the initial atmospheric mass fractions $f_{\rm at}^0\,=\,M_{\rm at}^0/M_{\rm pl}^0$, we employ the values predicted by the analytical approximation of \citet{mordasini2020} for planets of the given temperature orbiting a Sun-like star. Table~\ref{tab:evo_planets} gives the full list of considered planets. For illustration purposes, we include an extra model identical to model 23, but with an initial atmospheric mass fraction twice as large as that predicted by \citet{mordasini2020}. To model the evolution of the host stars, we employ the stellar evolution code Mors \citep[][]{johnstone2021mors,spada2013} and assume that the stars are moderate rotators, namely we set the rotational periods of all stars to be 6\,days at the age of 150\,Myr. 
\begin{table}
    \centering
    \caption{Parameters of the planets employed to test the impact of different mass loss prescriptions on atmospheric evolution.}
    \begin{tabular}{cccccc}
        \hline
        \hline
        $N_{\rm sim}$ & $M_*$ & $\Teq^*$ & $a$ & $M_{\rm pl}^0$ & $f_{\rm at}^0$ \\
                    & [\Msun] & [K] & [AU] & [\Mer] &  \\
         \hline
        1 & 0.9 & 500 & 0.2689 & 5 & 0.064\\
        2 & 0.9 & 500 & 0.2689 & 9 & 0.116\\
        3 & 0.9 & 500 & 0.2689 & 17 & 0.200\\
        4 & 0.9 & 500 & 0.2689 & 30 & 0.300\\
        5 & 0.9 & 1000 & 0.0672 & 5 & 0.026\\
        6 & 0.9 & 1000 & 0.0672 & 9 & 0.050\\
        7 & 0.9 & 1000 & 0.0672 & 17 & 0.098\\
        8 & 0.9 & 1000 & 0.0672 & 30 & 0.165\\
        9  & 0.9 & 1700 & 0.0233 & 5 & 0.010\\
        10 & 0.9 & 1700 & 0.0233 & 9 & 0.025\\
        11 & 0.9 & 1700 & 0.0233 & 17 & 0.050\\
        12 & 0.9 & 1700 & 0.0233 & 30 & 0.090\\        
        13 & 0.7 & 1700 & 0.0113 & 5 & 0.010\\
        14 & 0.7 & 1700 & 0.0113 & 9 & 0.025\\
        15 & 0.7 & 1700 & 0.0113 & 17 & 0.050\\
        16 & 0.7 & 1700 & 0.0113 & 30 & 0.090\\
        17 & 0.5 & 500 & 0.0603 & 5 & 0.064\\
        18 & 0.5 & 500 & 0.0603 & 9 & 0.116\\
        19 & 0.5 & 500 & 0.0603 & 17 & 0.200\\
        20 & 0.5 & 500 & 0.0603 & 30 & 0.300\\
        21 & 0.5 & 1000 & 0.0151 & 5 & 0.026\\
        22 & 0.5 & 1000 & 0.0151 & 9 & 0.050\\
        23 & 0.5 & 1000 & 0.0151 & 17 & 0.098\\
        24 & 0.5 & 1000 & 0.0151 & 30 & 0.165\\
        25 & 0.5 & 1000 & 0.0151 & 17 & 0.196\\
        \hline
    \end{tabular}
    \footnotesize{{\\$^*$ \Teq\ values are given as an average value as the actual values of \Teq\ vary across planetary evolution according to the stellar $L_{\rm bol}$ input.}}
    \label{tab:evo_planets}
\end{table}

\begin{figure*}
    \centering
    \includegraphics[width=1.0\linewidth]{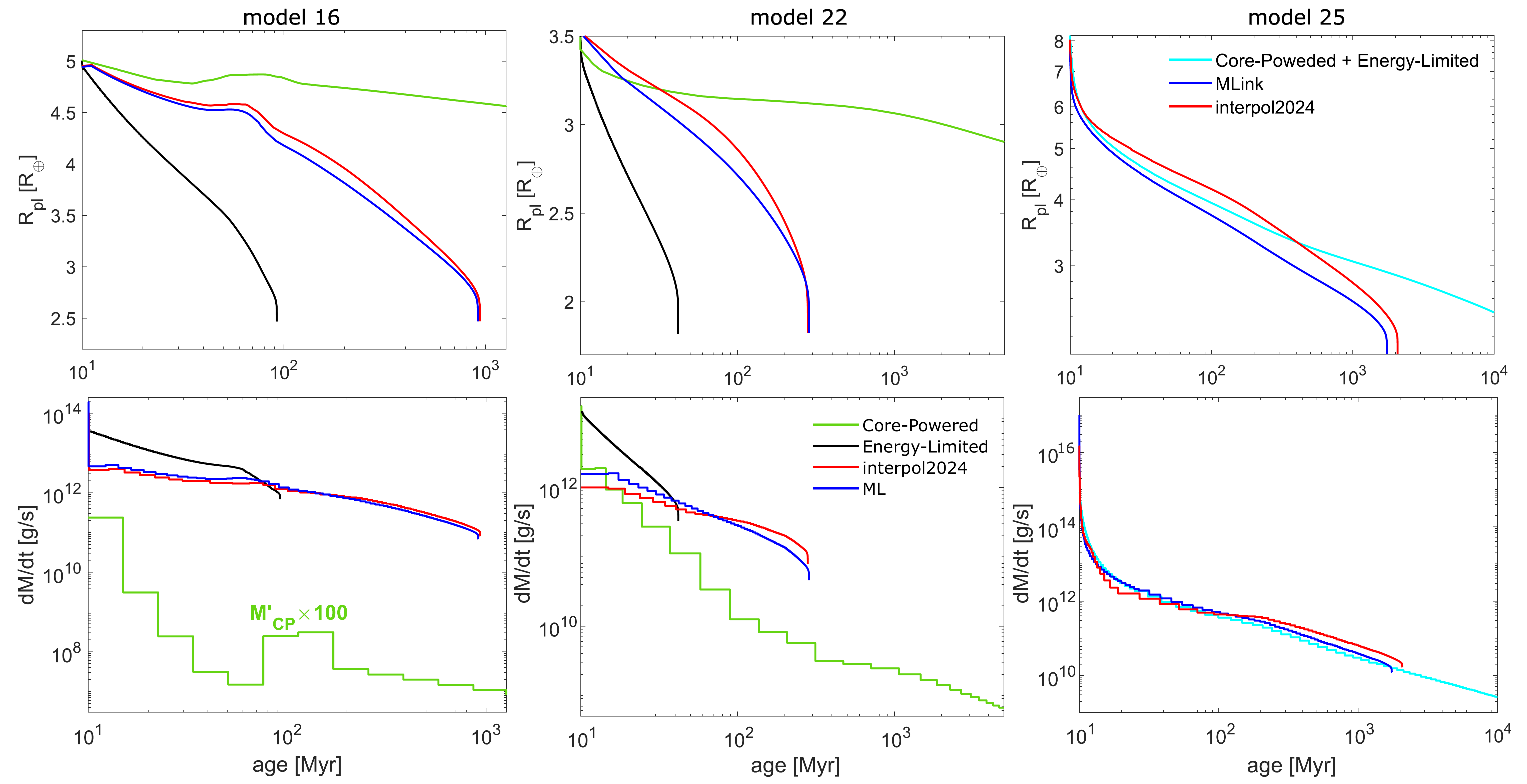}
    \caption{Planetary radius (top) and mass-loss rate (bottom) evolutionary tracks for planets number 16 (left; $M_*$\,=\,0.7\,\Msun; \Teq\,=\,1700\,K; $a$\,=\,0.0113\,AU; $M_{\rm pl}^0$\,=\,30\,\Mer; $f_{\rm at}^0$\,=\,0.090), 22 (center; $M_*$\,=\,0.5\,\Msun; \Teq\,=\,1000\,K; $a$\,=\,0.0151\,AU; $M_{\rm pl}^0$\,=\,9\,\Mer; $f_{\rm at}^0$\,=\,0.050), and 25 (right; ; $M_*$\,=\,0.5\,\Msun; \Teq\,=\,1000\,K; $a$\,=\,0.0151\,AU; $M_{\rm pl}^0$\,=\,17\,\Mer; $f_{\rm at}^0$\,=\,0.196; see Table~\ref{tab:evo_planets}), as predicted employing \texttt{MLink} (blue lines), \texttt{interpol2024} (red lines), energy-limited approximation (black lines), and core-powered mass-loss (green lines), or the combination of the two latter (cyan).}
    \label{fig:evo_tracks}
\end{figure*}

For each simulation, we let the planet evolve for 10\,Gyr or until the atmosphere is fully evaporated (i.e. atmospheric mass fraction equal to zero). As an example, Figure~\ref{fig:evo_tracks} shows the evolutionary tracks of three highly irradiated planets: a 9\,\Mer\ planet orbiting an 0.5\,\Msun\ star at 0.0151\,AU (i.e. $\Teq\,\simeq\,1000$\,K; simulation 22 in Table~\ref{tab:evo_planets}), a 17\,\Mer\ planet orbiting an 0.5\,\Msun\ star at 0.0151\,AU and starting its evolution with $f_{at}^0$ set to be twice of what predicted by approximation given in \citet[][i.e. $\Teq\,\simeq\,1000$\,K; simulation 25 in Table~\ref{tab:evo_planets}]{mordasini2020}, and of a 30\,\Mer\ planet orbiting an 0.7\,\Msun\ star at 0.0113\,AU (i.e. $\Teq\,\simeq\,1700$\,K; simulation 16 in Table~\ref{tab:evo_planets}). For all planets, lines of different colours show the tracks predicted using different atmospheric escape models as detailed in the legend. 

In all cases, the \texttt{MLink} and \texttt{interpol2024} predict similar escape rates, and thus similar evolution of planetary radii. We have already shown in Section~\ref{sec:interpolation_challenges} that the energy-limited approximation can drastically underestimate the escape from low-mass puffy planets \citep[see also][which is the case for model 25]{krenn2021}. For more compact planets (in particular, highly irradiated ones with $\Feuv>10^4$\,\ergscm), it can instead overestimate the escape as it ignores radiative recombination and losses \citep[e.g.][this is the case for models 16 and 22]{mc2009,Kurokawa_Nakamoto2014ApJ...783...54K}. For all planets in Figure~\ref{fig:evo_tracks}, the EUV flux exceeds $10^5$\,\ergscm\ at the beginning of the evolution, and thus the energy-limited approximation leads to atmospheric lifetimes of $\sim$6 (for the 9\,\Mer\ planet, model 22) and $\sim$10 (for the 30\,\Mer\ planet, model 16) shorter compared to the hydrodynamic model (i.e. \texttt{MLink} and \texttt{interplol2024}). For both models 16 and 22, the atmosphere evaporates completely before $\sim$1\,Gyr, therefore this inaccuracy of the energy-limited prescription does not play a role in shaping the predicted mass-radius distribution of this kind of planets at the ages of a few billion years. However, it can be essential for adequately studying young planetary systems.

Instead, the core-powered mass loss approximation tends to underestimate atmospheric loss for compact planets on long timescales. Thus, for the 9\,\Mer\ planet in Figure~\ref{fig:evo_tracks}, the core-powered mass loss at the beginning of the simulation is about an order of magnitude above the prediction of both \texttt{MLink} and \texttt{interpol2024}. However, due to the planet contraction and decline of \Teq, the strong escape ceases within a few tens of Myr and $\dot{M}_{\rm CP}$ does not exceed $10^{10}$\,g\,s$^{-1}$ after 100\,Myr. Therefore, the planet preserves its atmosphere until the end of the simulation. For the 30\,\Mer\ planet, despite its higher \Teq, core-powered mass loss remains negligible ($\sim$1000 times lower than that predicted by hydrodynamic modelling) throughout the evolution and the final radius resembles the one that would be obtained without considering atmospheric escape. Finally, the right panel of Figure~\ref{fig:evo_tracks} shows that combining energy-limited and core-powered mass-loss rates also does not necessarily lead to reproduce what predicted by hydrodynamic models.

\begin{figure}
    \centering
    \includegraphics[width=1\linewidth]{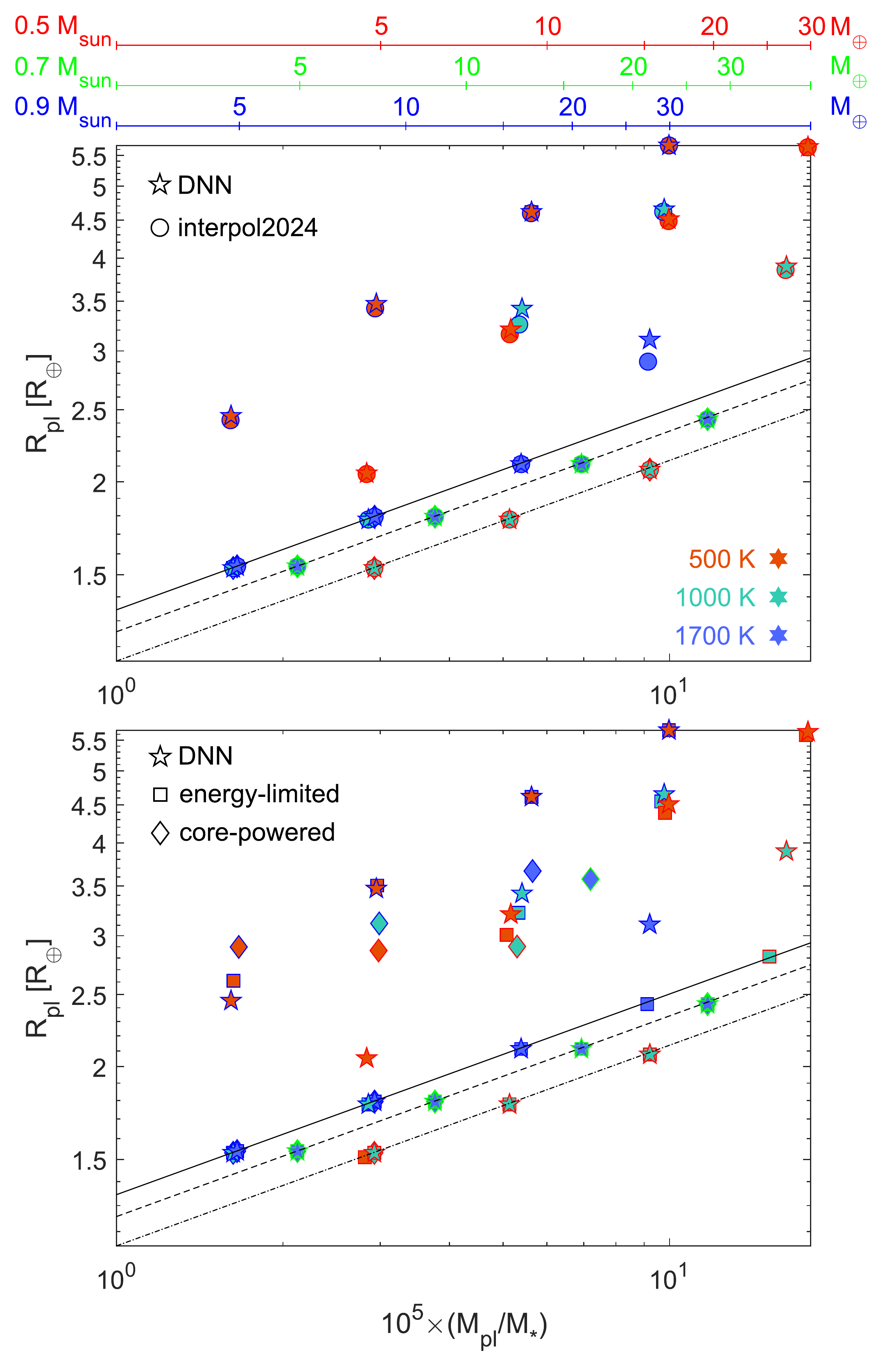}
    \caption{Mass-radius distribution at 5\,Gyr as predicted by the evolution model for synthetic planets 1--24 employing different atmospheric mass-loss models. Top: comparison of \texttt{MLink} (stars) with \texttt{interpol2024} (circles). Bottom panel: comparison of \texttt{MLink} with energy-limited approximation (squares) and core-powered mass loss (diamonds). In both panels, the symbol fill colour indicates the equilibrium temperature of the planet (red for 500\,K, turquoise for 1000\,K, and blue for 1700\,K), and the contour line of the symbols indicates the mass of the host star (blue line for 0.9\,\Msun, green for 0.7\,\Msun, and red for 0.5\,\Msun). To ease the comparison, the masses of the planets are normalised to the mass of their host stars, but the three x-axes at the top give the scale in Earth masses for each considered stellar mass. The three black lines, show the radius corresponding to the bare rocky core as a function of the host star mass (solid for 0.9\,\Msun, dashed for 0.7\,\Msun, and dash-dotted for 0.5\,\Msun).}
    \label{fig:evo_MR}
\end{figure}

In the top panel of Figure~\ref{fig:evo_MR}, we examine the performance of the \texttt{MLink} and \texttt{interpol2024} interpolations by comparing the positions of the model planets 1--24 in the mass-radius diagram at the age of 5\,Gyr. For better visualisation of the results, we plot the radii against the planetary masses normalised to the masses of their hosts. One can see that, despite some localised differences in the escape rates, the two methods predict, in general, very similar planetary parameters at ages of billions of years. However, for some planets, the difference in the predicted radii exceeds the typical observational uncertainties of $\sim$5\%, and remarkably these are the planets located close to the boundary between planets that preserve their primordial atmosphere and those that lose it (i.e. planets close to the radius gap; the largest difference among our test planets is achieved for models 7 and 12). This highlights the importance of accurate mass loss determination in studying phenomena such as the radius gap and the hot Neptune desert.

The bottom panel of Figure~\ref{fig:evo_MR} matches the predictions of the \texttt{MLink} with those of the analytical approximations. The energy-limited approximation (in the formulation described in Appendix~\ref{apx:EL-CP-RLO}) is accurate for planets under moderate conditions, but fails for planets near the radius gap (see e.g. model planets 12, 17, and 24). For the core-powered mass loss, for which we only show the models where the escape rates exceed $10^5$\,g\,s$^{-1}$, we find that it tends to overestimate the planetary radii, particularly for planets near the radius gap. 

\begin{figure}
    \centering
    \includegraphics[width=1\linewidth]{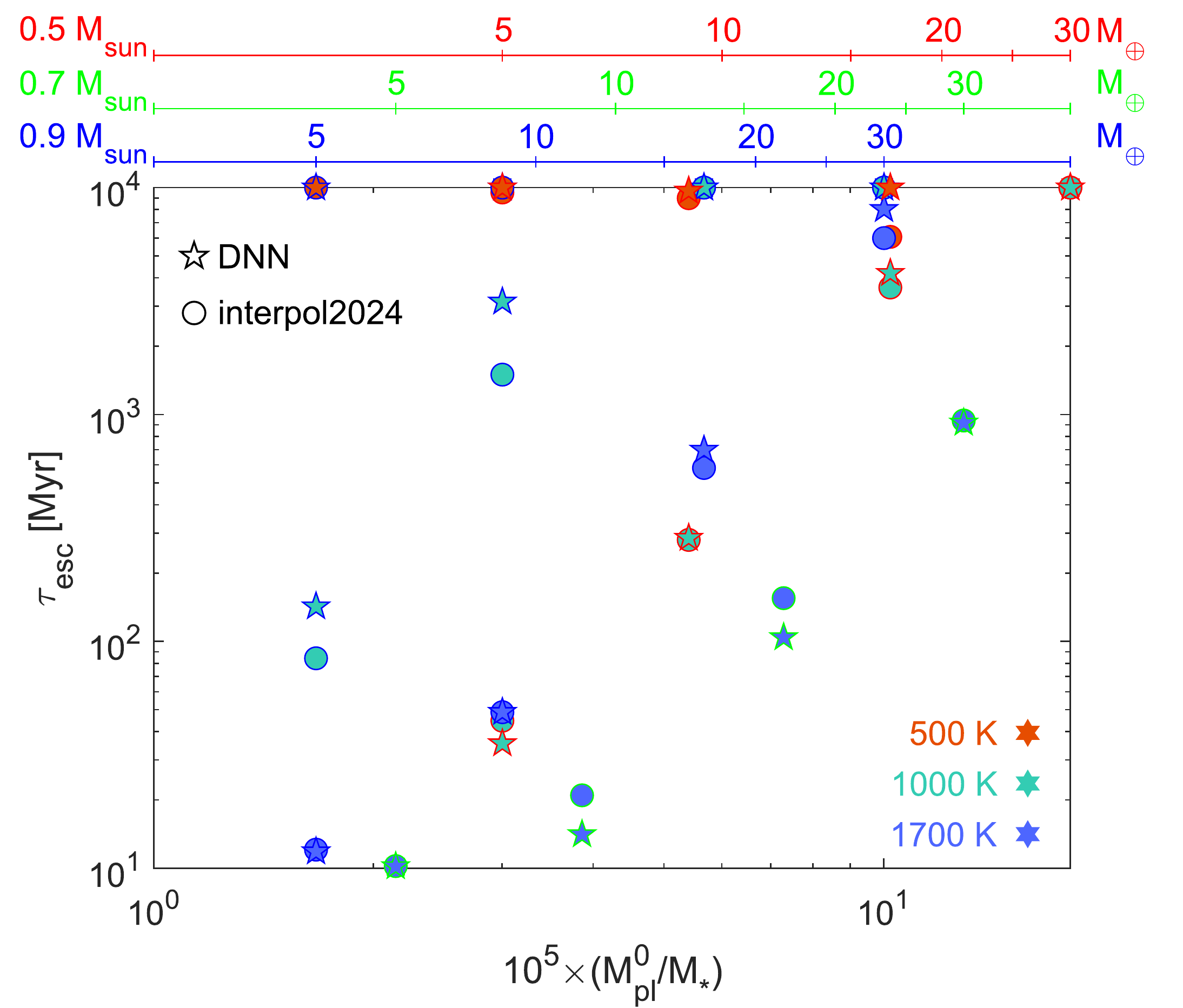}
    \caption{Comparison of the atmosphere evaporation time $\tau_{\rm esc}$ between the evolutionary models employing \texttt{MLink} (stars) and \texttt{interpol2024} (circles). $\tau_{\rm esc}$ is shown against the planetary initial mass normalised to the host star mass, but the three x-axes at the top give the scale in Earth masses for each considered stellar mass. The symbol fill colour and contour are the same as in Figure~\ref{fig:evo_MR}.}
    \label{fig:evo_tau_esc}
\end{figure}

Finally, Figure~\ref{fig:evo_tau_esc} compares the atmospheric evaporation timescales between the \texttt{MLink} and \texttt{interpol2024} for planets that fail to preserve their primordial atmospheres. We find comparable results between the two interpolation methods and differences become noticeable for planets with $\Teq\sim1000$\,K orbiting stars of 0.7\,\Msun\ and 0.9\,\Msun\ (i.e. planets losing their atmospheres at a moderate pace).

\section{Summary and concluding remarks}\label{sec:conclusions}%
The mass-loss rates of planets with hydrodynamic atmospheres depend on many parameters in a non-linear way and their values can span as much as ten orders of magnitude. Therefore, estimating the atmospheric mass-loss rates for a given planet is a non-trivial task that in general requires the use of elaborated hydrodynamic models at large computational costs, precluding their use in planet population and evolution studies, and thus forcing one to employ (over-)simplified analytical approximations. This problem can be overcome by employing a large grid of pre-calculated upper atmosphere models, further interpolating among them \citep{kubyshkina2018grid,kubyshkina2018app,kubyshkina_fossati2021}. However, accurate interpolation within the grid remains a challenging task.

The interpolation scheme developed in our earlier works \citep[see][]{kubyshkina2018grid,kubyshkina_fossati2021} has allowed for a significant improvement over commonly employed analytical approximations \citep[e.g.][]{krenn2021}. However, it is still subject to significant interpolation errors in some specific regions of the parameter space. To overcome this problem, we introduced a new interpolation approach, called \texttt{MLink}, based on dense neural network machine learning scheme to estimate mass-loss rates from the existing grid of hydrodynamic upper atmosphere models. This new approach led to a more reliable interpolation, which has improved the quality of the mass-loss estimates over previous methods.

We test the new interpolation method by studying the evolution of a small sample of planets with parameters chosen in such a way to maximise the impact of interpolation errors. To model atmospheric evolution, we employed the framework developed in \citet{kubyshkina2020mesa,kubyshkina2022MR} and performed the modelling for a range of Neptune-like planets with masses in the 5--30\,\Mer\ range in different orbits and around different stellar hosts. We found that for most cases, \texttt{MLink} and the old mass loss interpolation schemes lead to comparable planetary parameters at Gyr-timescales, though the evolutionary paths can be slightly different during some parts of the evolution. Furthermore, we find that for some planets, particularly for those at the top edge of the radius gap, the difference between the predicted planetary radii at a given age of tracks obtained with \texttt{MLink} and classical interpolation schemes can exceed the typical observational uncertainties.

Our results clearly indicate that machine learning can be successfully used to estimate atmospheric mass-loss rates from a model grid. This paves the way to the exploration of future larger and more complex grids of models that will have to be computed accounting for more physical processes, such as different atmospheric compositions or the impact of various photospheric conditions at the lower boundary (e.g. presence of aerosols).

\begin{acknowledgements}
D.K. was supported by a Schr\"odinger Fellowship supported by the Austrian Science Fund (FWF) project number J4792 (FEPLowS). We thank the anonymous referee for their comments that helped improving the presentation of the results.
\end{acknowledgements}

\bibliographystyle{aa} 
\bibliography{interpolator3} 

\begin{thebibliography}{86}
\expandafter\ifx\csname natexlab\endcsname\relax\def\natexlab#1{#1}\fi

\bibitem[{Abiodun {et~al.}(2018)Abiodun, Jantan, Omolara, Dada, Mohamed, \& Arshad}]{abiodun2018state}
Abiodun, O.~I., Jantan, A., Omolara, A.~E., {et~al.} 2018, Heliyon, 4

\bibitem[{{Affolter} {et~al.}(2023){Affolter}, {Mordasini}, {Oza}, {Kubyshkina}, \& {Fossati}}]{affolter2023}
{Affolter}, L., {Mordasini}, C., {Oza}, A.~V., {Kubyshkina}, D., \& {Fossati}, L. 2023, \aap, 676, A119

\bibitem[{Albawi {et~al.}(2017)Albawi, Mohammed, \& Al-Zawi}]{albawi2017understanding}
Albawi, S., Mohammed, T.~A., \& Al-Zawi, S. 2017, in 2017 international conference on engineering and technology (ICET), Ieee, 1--6

\bibitem[{{Benz} {et~al.}(2021){Benz}, {Broeg}, {Fortier}, {Rando}, {Beck}, {Beck}, {Queloz}, {Ehrenreich}, {Maxted}, {Isaak}, {Billot}, {Alibert}, {Alonso}, {Ant{\'o}nio}, {Asquier}, {Bandy}, {B{\'a}rczy}, {Barrado}, {Barros}, {Baumjohann}, {Bekkelien}, {Bergomi}, {Biondi}, {Bonfils}, {Borsato}, {Brandeker}, {Busch}, {Cabrera}, {Cessa}, {Charnoz}, {Chazelas}, {Collier Cameron}, {Corral Van Damme}, {Cortes}, {Davies}, {Deleuil}, {Deline}, {Delrez}, {Demangeon}, {Demory}, {Erikson}, {Farinato}, {Fossati}, {Fridlund}, {Futyan}, {Gandolfi}, {Garcia Munoz}, {Gillon}, {Guterman}, {Gutierrez}, {Hasiba}, {Heng}, {Hernandez}, {Hoyer}, {Kiss}, {Kovacs}, {Kuntzer}, {Laskar}, {Lecavelier des Etangs}, {Lendl}, {L{\'o}pez}, {Lora}, {Lovis}, {L{\"u}ftinger}, {Magrin}, {Malvasio}, {Marafatto}, {Michaelis}, {de Miguel}, {Modrego}, {Munari}, {Nascimbeni}, {Olofsson}, {Ottacher}, {Ottensamer}, {Pagano}, {Palacios}, {Pall{\'e}}, {Peter}, {Piazza}, {Piotto}, {Pizarro}, {Pollaco}, {Ragazzoni}, {Ratti}, {Rauer}, {Ribas}, {Rieder},
  {Rohlfs}, {Safa}, {Salatti}, {Santos}, {Scandariato}, {S{\'e}gransan}, {Simon}, {Smith}, {Sordet}, {Sousa}, {Steller}, {Szab{\'o}}, {Szoke}, {Thomas}, {Tschentscher}, {Udry}, {Van Grootel}, {Viotto}, {Walter}, {Walton}, {Wildi}, \& {Wolter}}]{benz2021_cheops}
{Benz}, W., {Broeg}, C., {Fortier}, A., {et~al.} 2021, Experimental Astronomy, 51, 109

\bibitem[{{Bonfanti} {et~al.}(2024){Bonfanti}, {Brady}, {Wilson}, {Venturini}, {Egger}, {Brandeker}, {Sousa}, {Lendl}, {Simon}, {Queloz}, {Olofsson}, {Adibekyan}, {Alibert}, {Fossati}, {Hooton}, {Kubyshkina}, {Luque}, {Murgas}, {Mustill}, {Santos}, {Van Grootel}, {Alonso}, {Asquier}, {Bandy}, {B{\'a}rczy}, {Barrado Navascues}, {Barros}, {Baumjohann}, {Bean}, {Beck}, {Beck}, {Benz}, {Bergomi}, {Billot}, {Borsato}, {Broeg}, {Collier Cameron}, {Csizmadia}, {Cubillos}, {Davies}, {Deleuil}, {Deline}, {Delrez}, {Demangeon}, {Demory}, {Ehrenreich}, {Erikson}, {Fortier}, {Fridlund}, {Gandolfi}, {Gillon}, {G{\"u}del}, {G{\"u}nther}, {Heitzmann}, {Helling}, {Hoyer}, {Isaak}, {Kasper}, {Kiss}, {Lam}, {Laskar}, {Lecavelier des Etangs}, {Magrin}, {Maxted}, {Mordasini}, {Nascimbeni}, {Ottensamer}, {Pagano}, {Pall{\'e}}, {Peter}, {Piotto}, {Pollacco}, {Ragazzoni}, {Rando}, {Rauer}, {Ribas}, {Scandariato}, {S{\'e}gransan}, {Seifahrt}, {Smith}, {Stalport}, {Stef{\'a}nsson}, {Steinberger}, {St{\"u}rmer}, {Szab{\'o}}, {Thomas},
  {Udry}, {Villaver}, {Walton}, {Westerdorff}, \& {Zingales}}]{bonfanti2023_toi732}
{Bonfanti}, A., {Brady}, M., {Wilson}, T.~G., {et~al.} 2024, \aap, 682, A66

\bibitem[{{Bonfanti} {et~al.}(2021){Bonfanti}, {Fossati}, {Kubyshkina}, \& {Cubillos}}]{bonfanti2021evol}
{Bonfanti}, A., {Fossati}, L., {Kubyshkina}, D., \& {Cubillos}, P.~E. 2021, \aap, 656, A157

\bibitem[{Buhmann(2000)}]{buhmann2000radial}
Buhmann, M.~D. 2000, Acta numerica, 9, 1

\bibitem[{{Carleo} {et~al.}(2020){Carleo}, {Gandolfi}, {Barrag{\'a}n}, {Livingston}, {Persson}, {Lam}, {Vidotto}, {Lund}, {Villarreal D'Angelo}, {Collins}, {Fossati}, {Howard}, {Kubyshkina}, {Brahm}, {Oklop{\v{c}}i{\'c}}, {Molli{\`e}re}, {Redfield}, {Serrano}, {Dai}, {Fridlund}, {Borsa}, {Korth}, {Esposito}, {D{\'\i}az}, {Dyregaard Nielsen}, {Hellier}, {Mathur}, {Deeg}, {Hatzes}, {Benatti}, {Rodler}, {Alarcon}, {Spina}, {Santos}, {Georgieva}, {Garc{\'\i}a}, {Gonz{\'a}lez-Cuesta}, {Ricker}, {Vanderspek}, {Latham}, {Seager}, {Winn}, {Jenkins}, {Albrecht}, {Batalha}, {Beard}, {Boyd}, {Bouchy}, {Burt}, {Butler}, {Cabrera}, {Chontos}, {Ciardi}, {Cochran}, {Collins}, {Crane}, {Crossfield}, {Csizmadia}, {Dragomir}, {Dressing}, {Eigm{\"u}ller}, {Endl}, {Erikson}, {Espinoza}, {Fausnaugh}, {Feng}, {Flowers}, {Fulton}, {Gonzales}, {Grieves}, {Grziwa}, {Guenther}, {Guerrero}, {Henning}, {Hidalgo}, {Hirano}, {Hjorth}, {Huber}, {Isaacson}, {Jones}, {Jord{\'a}n}, {Kab{\'a}th}, {Kane}, {Knudstrup}, {Lubin}, {Luque},
  {Mireles}, {Narita}, {Nespral}, {Niraula}, {Nowak}, {Palle}, {P{\"a}tzold}, {Petigura}, {Prieto-Arranz}, {Rauer}, {Robertson}, {Rose}, {Roy}, {Sarkis}, {Schlieder}, {S{\'e}gransan}, {Shectman}, {Skarka}, {Smith}, {Smith}, {Stassun}, {Teske}, {Twicken}, {Van Eylen}, {Wang}, {Weiss}, \& {Wyttenbach}}]{carleo2020}
{Carleo}, I., {Gandolfi}, D., {Barrag{\'a}n}, O., {et~al.} 2020, \aj, 160, 114

\bibitem[{{Chen} \& {Rogers}(2016)}]{chen_rogers2016}
{Chen}, H. \& {Rogers}, L.~A. 2016, \apj, 831, 180

\bibitem[{{Choi} {et~al.}(2016){Choi}, {Dotter}, {Conroy}, {Cantiello}, {Paxton}, \& {Johnson}}]{choi2016}
{Choi}, J., {Dotter}, A., {Conroy}, C., {et~al.} 2016, \apj, 823, 102

\bibitem[{{Cubillos} {et~al.}(2017){Cubillos}, {Fossati}, {Erkaev}, {Malik}, {Tokano}, {Lendl}, {Johnstone}, {Lammer}, \& {Wyttenbach}}]{cubillos2017}
{Cubillos}, P.~E., {Fossati}, L., {Erkaev}, N.~V., {et~al.} 2017, \apj, 849, 145

\bibitem[{{Erkaev} {et~al.}(2007){Erkaev}, {Kulikov}, {Lammer}, {Selsis}, {Langmayr}, {Jaritz}, \& {Biernat}}]{Erkaev2007}
{Erkaev}, N.~V., {Kulikov}, Y.~N., {Lammer}, H., {et~al.} 2007, \aap, 472, 329

\bibitem[{{Fossati} {et~al.}(2017){Fossati}, {Erkaev}, {Lammer}, {Cubillos}, {Odert}, {Juvan}, {Kislyakova}, {Lendl}, {Kubyshkina}, \& {Bauer}}]{fossati2017}
{Fossati}, L., {Erkaev}, N.~V., {Lammer}, H., {et~al.} 2017, \aap, 598, A90

\bibitem[{{Freedman} {et~al.}(2008){Freedman}, {Marley}, \& {Lodders}}]{freedman2008}
{Freedman}, R.~S., {Marley}, M.~S., \& {Lodders}, K. 2008, \apjs, 174, 504

\bibitem[{{Gandolfi} {et~al.}(2019){Gandolfi}, {Fossati}, {Livingston}, {Stassun}, {Grziwa}, {Barrag{\'a}n}, {Fridlund}, {Kubyshkina}, {Persson}, {Dai}, {Lam}, {Albrecht}, {Batalha}, {Beck}, {Justesen}, {Cabrera}, {Cartwright}, {Cochran}, {Csizmadia}, {Davies}, {Deeg}, {Eigm{\"u}ller}, {Endl}, {Erikson}, {Esposito}, {Garc{\'\i}a}, {Goeke}, {Gonz{\'a}lez-Cuesta}, {Guenther}, {Hatzes}, {Hidalgo}, {Hirano}, {Hjorth}, {Kabath}, {Knudstrup}, {Korth}, {Li}, {Luque}, {Mathur}, {Monta{\~n}es Rodr{\'\i}guez}, {Narita}, {Nespral}, {Niraula}, {Nowak}, {Palle}, {P{\"a}tzold}, {Prieto-Arranz}, {Rauer}, {Redfield}, {Ribas}, {Skarka}, {Smith}, {Rowden}, {Torres}, {Van Eylen}, \& {Vezie}}]{gandolfi2019_hd15337}
{Gandolfi}, D., {Fossati}, L., {Livingston}, J.~H., {et~al.} 2019, \apjl, 876, L24

\bibitem[{{Ginzburg} \& {Sari}(2016)}]{Ginzburg2016}
{Ginzburg}, S. \& {Sari}, R. 2016, \apj, 819, 116

\bibitem[{{Ginzburg} {et~al.}(2018){Ginzburg}, {Schlichting}, \& {Sari}}]{ginzburg2018}
{Ginzburg}, S., {Schlichting}, H.~E., \& {Sari}, R. 2018, \mnras, 476, 759

\bibitem[{Gu {et~al.}(2018)Gu, Wang, Kuen, Ma, Shahroudy, Shuai, Liu, Wang, Wang, Cai, {et~al.}}]{gu2018recent}
Gu, J., Wang, Z., Kuen, J., {et~al.} 2018, Pattern recognition, 77, 354

\bibitem[{{G{\"u}nther} {et~al.}(2019){G{\"u}nther}, {Pozuelos}, {Dittmann}, {Dragomir}, {Kane}, {Daylan}, {Feinstein}, {Huang}, {Morton}, {Bonfanti}, {Bouma}, {Burt}, {Collins}, {Lissauer}, {Matthews}, {Montet}, {Vanderburg}, {Wang}, {Winters}, {Ricker}, {Vanderspek}, {Latham}, {Seager}, {Winn}, {Jenkins}, {Armstrong}, {Barkaoui}, {Batalha}, {Bean}, {Caldwell}, {Ciardi}, {Collins}, {Crossfield}, {Fausnaugh}, {Furesz}, {Gan}, {Gillon}, {Guerrero}, {Horne}, {Howell}, {Ireland}, {Isopi}, {Jehin}, {Kielkopf}, {Lepine}, {Mallia}, {Matson}, {Myers}, {Palle}, {Quinn}, {Relles}, {Rojas-Ayala}, {Schlieder}, {Sefako}, {Shporer}, {Su{\'a}rez}, {Tan}, {Ting}, {Twicken}, \& {Waite}}]{gunther2019_toi270}
{G{\"u}nther}, M.~N., {Pozuelos}, F.~J., {Dittmann}, J.~A., {et~al.} 2019, Nature Astronomy, 3, 1099

\bibitem[{{Guo}(2024)}]{Guo2024NatAs.tmp...89G}
{Guo}, J.~H. 2024, Nature Astronomy, 8, 920

\bibitem[{{Guo} \& {Ben-Jaffel}(2016)}]{guo_benjaffel2016}
{Guo}, J.~H. \& {Ben-Jaffel}, L. 2016, \apj, 818, 107

\bibitem[{{Gupta} \& {Schlichting}(2019)}]{gupta_schlichting2019}
{Gupta}, A. \& {Schlichting}, H.~E. 2019, \mnras, 487, 24

\bibitem[{Gurney(2018)}]{gurney2018introduction}
Gurney, K. 2018, An introduction to neural networks (CRC press)

\bibitem[{{Hadden} \& {Lithwick}(2017)}]{hadden2017}
{Hadden}, S. \& {Lithwick}, Y. 2017, \aj, 154, 5

\bibitem[{Hecht-Nielsen(1992)}]{hecht1992theory}
Hecht-Nielsen, R. 1992, in Neural networks for perception (Elsevier), 65--93

\bibitem[{Hinton(2012)}]{hinton2012improving}
Hinton, G. 2012, arXiv:1207.0580

\bibitem[{Hochreiter \& Schmidhuber(1997)}]{hochreiter1997long}
Hochreiter, S. \& Schmidhuber, J. 1997, Neural computation, 9, 1735

\bibitem[{{Jackson} {et~al.}(2012){Jackson}, {Davis}, \& {Wheatley}}]{jackson2012}
{Jackson}, A.~P., {Davis}, T.~A., \& {Wheatley}, P.~J. 2012, \mnras, 422, 2024

\bibitem[{{Jin} \& {Mordasini}(2018)}]{jin2018}
{Jin}, S. \& {Mordasini}, C. 2018, \apj, 853, 163

\bibitem[{{Johnstone} {et~al.}(2021){Johnstone}, {Bartel}, \& {G{\"u}del}}]{johnstone2021mors}
{Johnstone}, C.~P., {Bartel}, M., \& {G{\"u}del}, M. 2021, \aap, 649, A96

\bibitem[{{Ketzer} \& {Poppenhaeger}(2022)}]{ketzer2022}
{Ketzer}, L. \& {Poppenhaeger}, K. 2022, Astronomische Nachrichten, 343, e10105

\bibitem[{{Koskinen} {et~al.}(2022){Koskinen}, {Lavvas}, {Huang}, {Bergsten}, {Fernandes}, \& {Young}}]{Koskinen2022ApJ...929...52K}
{Koskinen}, T.~T., {Lavvas}, P., {Huang}, C., {et~al.} 2022, \apj, 929, 52

\bibitem[{{Krenn} {et~al.}(2021){Krenn}, {Fossati}, {Kubyshkina}, \& {Lammer}}]{krenn2021}
{Krenn}, A.~F., {Fossati}, L., {Kubyshkina}, D., \& {Lammer}, H. 2021, \aap, 650, A94

\bibitem[{{Kubyshkina}(2022)}]{kubyshkina2022AN....34310077K}
{Kubyshkina}, D. 2022, Astronomische Nachrichten, 343, e10077

\bibitem[{{Kubyshkina} \& {Fossati}(2022)}]{kubyshkina2022MR}
{Kubyshkina}, D. \& {Fossati}, L. 2022, \aap, 668, A178

\bibitem[{{Kubyshkina} {et~al.}(2024){Kubyshkina}, {Fossati}, \& {Erkaev}}]{kubyshkina2024cloudy}
{Kubyshkina}, D., {Fossati}, L., \& {Erkaev}, N.~V. 2024, \aap, 684, A26

\bibitem[{{Kubyshkina} {et~al.}(2018{\natexlab{a}}){Kubyshkina}, {Fossati}, {Erkaev}, {Cubillos}, {Johnstone}, {Kislyakova}, {Lammer}, {Lendl}, \& {Odert}}]{kubyshkina2018app}
{Kubyshkina}, D., {Fossati}, L., {Erkaev}, N.~V., {et~al.} 2018{\natexlab{a}}, \apjl, 866, L18

\bibitem[{{Kubyshkina} {et~al.}(2018{\natexlab{b}}){Kubyshkina}, {Fossati}, {Erkaev}, {Johnstone}, {Cubillos}, {Kislyakova}, {Lammer}, {Lendl}, \& {Odert}}]{kubyshkina2018grid}
{Kubyshkina}, D., {Fossati}, L., {Erkaev}, N.~V., {et~al.} 2018{\natexlab{b}}, \aap, 619, A151

\bibitem[{{Kubyshkina} {et~al.}(2019){Kubyshkina}, {Fossati}, {Mustill}, {Cubillos}, {Davies}, {Erkaev}, {Johnstone}, {Kislyakova}, {Lammer}, {Lendl}, \& {Odert}}]{kubyshkina2019kepler11}
{Kubyshkina}, D., {Fossati}, L., {Mustill}, A.~J., {et~al.} 2019, \aap, 632, A65

\bibitem[{{Kubyshkina} {et~al.}(2020){Kubyshkina}, {Vidotto}, {Fossati}, \& {Farrell}}]{kubyshkina2020mesa}
{Kubyshkina}, D., {Vidotto}, A.~A., {Fossati}, L., \& {Farrell}, E. 2020, \mnras, 499, 77

\bibitem[{{Kubyshkina} {et~al.}(2022){Kubyshkina}, {Vidotto}, {Villarreal D'Angelo}, {Carolan}, {Hazra}, \& {Carleo}}]{kubyshkina2022_3d}
{Kubyshkina}, D., {Vidotto}, A.~A., {Villarreal D'Angelo}, C., {et~al.} 2022, \mnras, 510, 2111

\bibitem[{{Kubyshkina} \& {Fossati}(2021)}]{kubyshkina_fossati2021}
{Kubyshkina}, D.~I. \& {Fossati}, L. 2021, Research Notes of the American Astronomical Society, 5, 74

\bibitem[{{Kurokawa} \& {Nakamoto}(2014)}]{Kurokawa_Nakamoto2014ApJ...783...54K}
{Kurokawa}, H. \& {Nakamoto}, T. 2014, \apj, 783, 54

\bibitem[{{Lacedelli} {et~al.}(2021){Lacedelli}, {Malavolta}, {Borsato}, {Piotto}, {Nardiello}, {Mortier}, {Stalport}, {Collier Cameron}, {Poretti}, {Buchhave}, {L{\'o}pez-Morales}, {Nascimbeni}, {Wilson}, {Udry}, {Latham}, {Bonomo}, {Damasso}, {Dumusque}, {Jenkins}, {Lovis}, {Rice}, {Sasselov}, {Winn}, {Andreuzzi}, {Cosentino}, {Charbonneau}, {Di Fabrizio}, {Martnez Fiorenzano}, {Ghedina}, {Harutyunyan}, {Lienhard}, {Micela}, {Molinari}, {Pagano}, {Pepe}, {Phillips}, {Pinamonti}, {Ricker}, {Scandariato}, {Sozzetti}, \& {Watson}}]{Lacedelli2021}
{Lacedelli}, G., {Malavolta}, L., {Borsato}, L., {et~al.} 2021, \mnras, 501, 4148

\bibitem[{{Lam} {et~al.}(2023){Lam}, {Cabrera}, {Hooton}, {Alibert}, {Bonfanti}, {Beck}, {Deline}, {Flor{\'e}n}, {Simon}, {Fossati}, {Persson}, {Fridlund}, {Salmon}, {Hoyer}, {Osborn}, {Wilson}, {Georgieva}, {Nowak}, {Luque}, {Egger}, {Adibekyan}, {Alonso}, {Escud{\'e}}, {B{\'a}rczy}, {Barrado}, {Barros}, {Baumjohann}, {Beck}, {Bekkelien}, {Benz}, {Billot}, {Bonfils}, {Brandeker}, {Broeg}, {Charnoz}, {Cameron}, {Csizmadia}, {Davies}, {Deleuil}, {Delrez}, {Demangeon}, {Demory}, {Ehrenreich}, {Erikson}, {Fortier}, {Futyan}, {Gandolfi}, {Gillon}, {Guedel}, {Guterman}, {Laskar}, {Latham}, {Lecavelier des Etangs}, {Lendl}, {Lovis}, {Heng}, {Isaak}, {Kiss}, {Magrin}, {Maxted}, {Nascimbeni}, {Olofsson}, {Ottensamer}, {Pagano}, {Pall{\'e}}, {Peter}, {Piotto}, {Pollacco}, {Queloz}, {Ribas}, {Ragazzoni}, {Rando}, {Rauer}, {Santos}, {Scandariato}, {Seager}, {S{\'e}gransan}, {Serrano}, {Smith}, {Sousa}, {Steller}, {Szab{\'o}}, {Thomas}, {Udry}, {Van Grootel}, {Walton}, \& {Winn}}]{lam2023_toi1260}
{Lam}, K.~W.~F., {Cabrera}, J., {Hooton}, M.~J., {et~al.} 2023, \mnras, 519, 1437

\bibitem[{{Lee} \& {Chiang}(2015)}]{lee_chiang2015ApJ...811...41L}
{Lee}, E.~J. \& {Chiang}, E. 2015, \apj, 811, 41

\bibitem[{{Lee} \& {Chiang}(2017)}]{lee_chiang2017}
{Lee}, E.~J. \& {Chiang}, E. 2017, \apj, 842, 40

\bibitem[{{Lee} {et~al.}(2022){Lee}, {Karalis}, \& {Thorngren}}]{evelee2022fgap}
{Lee}, E.~J., {Karalis}, A., \& {Thorngren}, D.~P. 2022, \apj, 941, 186

\bibitem[{{Leleu} {et~al.}(2021){Leleu}, {Alibert}, {Hara}, {Hooton}, {Wilson}, {Robutel}, {Delisle}, {Laskar}, {Hoyer}, {Lovis}, {Bryant}, {Ducrot}, {Cabrera}, {Delrez}, {Acton}, {Adibekyan}, {Allart}, {Allende Prieto}, {Alonso}, {Alves}, {Anderson}, {Angerhausen}, {Anglada Escud{\'e}}, {Asquier}, {Barrado}, {Barros}, {Baumjohann}, {Bayliss}, {Beck}, {Beck}, {Bekkelien}, {Benz}, {Billot}, {Bonfanti}, {Bonfils}, {Bouchy}, {Bourrier}, {Bou{\'e}}, {Brandeker}, {Broeg}, {Buder}, {Burdanov}, {Burleigh}, {B{\'a}rczy}, {Cameron}, {Chamberlain}, {Charnoz}, {Cooke}, {Corral Van Damme}, {Correia}, {Cristiani}, {Damasso}, {Davies}, {Deleuil}, {Demangeon}, {Demory}, {Di Marcantonio}, {Di Persio}, {Dumusque}, {Ehrenreich}, {Erikson}, {Figueira}, {Fortier}, {Fossati}, {Fridlund}, {Futyan}, {Gandolfi}, {Garc{\'\i}a Mu{\~n}oz}, {Garcia}, {Gill}, {Gillen}, {Gillon}, {Goad}, {Gonz{\'a}lez Hern{\'a}ndez}, {Guedel}, {G{\"u}nther}, {Haldemann}, {Henderson}, {Heng}, {Hogan}, {Isaak}, {Jehin}, {Jenkins}, {Jord{\'a}n}, {Kiss},
  {Kristiansen}, {Lam}, {Lavie}, {Lecavelier des Etangs}, {Lendl}, {Lillo-Box}, {Lo Curto}, {Magrin}, {Martins}, {Maxted}, {McCormac}, {Mehner}, {Micela}, {Molaro}, {Moyano}, {Murray}, {Nascimbeni}, {Nunes}, {Olofsson}, {Osborn}, {Oshagh}, {Ottensamer}, {Pagano}, {Pall{\'e}}, {Pedersen}, {Pepe}, {Persson}, {Peter}, {Piotto}, {Polenta}, {Pollacco}, {Poretti}, {Pozuelos}, {Queloz}, {Ragazzoni}, {Rando}, {Ratti}, {Rauer}, {Raynard}, {Rebolo}, {Reimers}, {Ribas}, {Santos}, {Scandariato}, {Schneider}, {Sebastian}, {Sestovic}, {Simon}, {Smith}, {Sousa}, {Sozzetti}, {Steller}, {Su{\'a}rez Mascare{\~n}o}, {Szab{\'o}}, {S{\'e}gransan}, {Thomas}, {Thompson}, {Tilbrook}, {Triaud}, {Turner}, {Udry}, {Van Grootel}, {Venus}, {Verrecchia}, {Vines}, {Walton}, {West}, {Wheatley}, {Wolter}, \& {Zapatero Osorio}}]{leleu2021_toi178}
{Leleu}, A., {Alibert}, Y., {Hara}, N.~C., {et~al.} 2021, \aap, 649, A26

\bibitem[{{Libby-Roberts} {et~al.}(2020){Libby-Roberts}, {Berta-Thompson}, {D{\'e}sert}, {Masuda}, {Morley}, {Lopez}, {Deck}, {Fabrycky}, {Fortney}, {Line}, {Sanchis-Ojeda}, \& {Winn}}]{Libby-Roberts2020}
{Libby-Roberts}, J.~E., {Berta-Thompson}, Z.~K., {D{\'e}sert}, J.-M., {et~al.} 2020, \aj, 159, 57

\bibitem[{{Luque} {et~al.}(2023){Luque}, {Osborn}, {Leleu}, {Pall{\'e}}, {Bonfanti}, {Barrag{\'a}n}, {Wilson}, {Broeg}, {Cameron}, {Lendl}, {Maxted}, {Alibert}, {Gandolfi}, {Delisle}, {Hooton}, {Egger}, {Nowak}, {Lafarga}, {Rapetti}, {Twicken}, {Morales}, {Carleo}, {Orell-Miquel}, {Adibekyan}, {Alonso}, {Alqasim}, {Amado}, {Anderson}, {Anglada-Escud{\'e}}, {Bandy}, {B{\'a}rczy}, {Barrado Navascues}, {Barros}, {Baumjohann}, {Bayliss}, {Bean}, {Beck}, {Beck}, {Benz}, {Billot}, {Bonfils}, {Borsato}, {Boyle}, {Brandeker}, {Bryant}, {Cabrera}, {Carrazco-Gaxiola}, {Charbonneau}, {Charnoz}, {Ciardi}, {Cochran}, {Collins}, {Crossfield}, {Csizmadia}, {Cubillos}, {Dai}, {Davies}, {Deeg}, {Deleuil}, {Deline}, {Delrez}, {Demangeon}, {Demory}, {Ehrenreich}, {Erikson}, {Esparza-Borges}, {Falk}, {Fortier}, {Fossati}, {Fridlund}, {Fukui}, {Garcia-Mejia}, {Gill}, {Gillon}, {Goffo}, {G{\'o}mez Maqueo Chew}, {G{\"u}del}, {Guenther}, {G{\"u}nther}, {Hatzes}, {Helling}, {Hesse}, {Howell}, {Hoyer}, {Ikuta}, {Isaak}, {Jenkins},
  {Kagetani}, {Kiss}, {Kodama}, {Korth}, {Lam}, {Laskar}, {Latham}, {Lecavelier des Etangs}, {Leon}, {Livingston}, {Magrin}, {Matson}, {Matthews}, {Mordasini}, {Mori}, {Moyano}, {Munari}, {Murgas}, {Narita}, {Nascimbeni}, {Olofsson}, {Osborne}, {Ottensamer}, {Pagano}, {Parviainen}, {Peter}, {Piotto}, {Pollacco}, {Queloz}, {Quinn}, {Quirrenbach}, {Ragazzoni}, {Rando}, {Ratti}, {Rauer}, {Redfield}, {Ribas}, {Ricker}, {Rudat}, {Sabin}, {Salmon}, {Santos}, {Scandariato}, {Schanche}, {Schlieder}, {Seager}, {S{\'e}gransan}, {Shporer}, {Simon}, {Smith}, {Sousa}, {Stalport}, {Szab{\'o}}, {Thomas}, {Tuson}, {Udry}, {Vanderburg}, {Van Eylen}, {Van Grootel}, {Venturini}, {Walter}, {Walton}, {Watanabe}, {Winn}, \& {Zingales}}]{luque2023}
{Luque}, R., {Osborn}, H.~P., {Leleu}, A., {et~al.} 2023, \nat, 623, 932

\bibitem[{{Masuda}(2014)}]{masuda2014}
{Masuda}, K. 2014, \apj, 783, 53

\bibitem[{{Matt} {et~al.}(2015){Matt}, {Brun}, {Baraffe}, {Bouvier}, \& {Chabrier}}]{matt2015}
{Matt}, S.~P., {Brun}, A.~S., {Baraffe}, I., {Bouvier}, J., \& {Chabrier}, G. 2015, \apjl, 799, L23

\bibitem[{{Modirrousta-Galian} {et~al.}(2020){Modirrousta-Galian}, {Locci}, \& {Micela}}]{modirrousta2020}
{Modirrousta-Galian}, D., {Locci}, D., \& {Micela}, G. 2020, \apj, 891, 158

\bibitem[{{Mordasini}(2020)}]{mordasini2020}
{Mordasini}, C. 2020, \aap, 638, A52

\bibitem[{{Mullally} {et~al.}(2015){Mullally}, {Coughlin}, {Thompson}, {Rowe}, {Burke}, {Latham}, {Batalha}, {Bryson}, {Christiansen}, {Henze}, {Ofir}, {Quarles}, {Shporer}, {Van Eylen}, {Van Laerhoven}, {Shah}, {Wolfgang}, {Chaplin}, {Xie}, {Akeson}, {Argabright}, {Bachtell}, {Barclay}, {Borucki}, {Caldwell}, {Campbell}, {Catanzarite}, {Cochran}, {Duren}, {Fleming}, {Fraquelli}, {Girouard}, {Haas}, {He{\l}miniak}, {Howell}, {Huber}, {Larson}, {Gautier}, {Jenkins}, {Li}, {Lissauer}, {McArthur}, {Miller}, {Morris}, {Patil-Sabale}, {Plavchan}, {Putnam}, {Quintana}, {Ramirez}, {Silva Aguirre}, {Seader}, {Smith}, {Steffen}, {Stewart}, {Stober}, {Still}, {Tenenbaum}, {Troeltzsch}, {Twicken}, \& {Zamudio}}]{mullally2015}
{Mullally}, F., {Coughlin}, J.~L., {Thompson}, S.~E., {et~al.} 2015, \apjs, 217, 31

\bibitem[{{M{\"u}ller} {et~al.}(2024){M{\"u}ller}, {Baron}, {Helled}, {Bouchy}, \& {Parc}}]{mueller2024_MRrelation}
{M{\"u}ller}, S., {Baron}, J., {Helled}, R., {Bouchy}, F., \& {Parc}, L. 2024, \aap, 686, A296

\bibitem[{{Murray-Clay} {et~al.}(2009){Murray-Clay}, {Chiang}, \& {Murray}}]{mc2009}
{Murray-Clay}, R.~A., {Chiang}, E.~I., \& {Murray}, N. 2009, \apj, 693, 23

\bibitem[{Narayan(1997)}]{narayan1997generalized}
Narayan, S. 1997, Information sciences, 99, 69

\bibitem[{{Nielsen} {et~al.}(2020){Nielsen}, {Gandolfi}, {Armstrong}, {Jenkins}, {Fridlund}, {Santos}, {Dai}, {Adibekyan}, {Luque}, {Steffen}, {Esposito}, {Meru}, {Sabotta}, {Bolmont}, {Kossakowski}, {Otegi}, {Murgas}, {Stalport}, {Rodler}, {D{\'\i}az}, {Kurtovic}, {Ricker}, {Vanderspek}, {Latham}, {Seager}, {Winn}, {Jenkins}, {Allart}, {Almenara}, {Barrado}, {Barros}, {Bayliss}, {Berdi{\~n}as}, {Boisse}, {Bouchy}, {Boyd}, {Brown}, {Bryant}, {Burke}, {Cochran}, {Cooke}, {Demangeon}, {D{\'\i}az}, {Dittman}, {Dorn}, {Dumusque}, {Garc{\'\i}a}, {Gonz{\'a}lez-Cuesta}, {Grziwa}, {Georgieva}, {Guerrero}, {Hatzes}, {Helled}, {Henze}, {Hojjatpanah}, {Korth}, {Lam}, {Lillo-Box}, {Lopez}, {Livingston}, {Mathur}, {Mousis}, {Narita}, {Osborn}, {Palle}, {Rojas}, {Persson}, {Quinn}, {Rauer}, {Redfield}, {Santerne}, {dos Santos}, {Seidel}, {Sousa}, {Ting}, {Turbet}, {Udry}, {Vanderburg}, {Van Eylen}, {Vines}, {Wheatley}, \& {Wilson}}]{nielsen2020_toi125}
{Nielsen}, L.~D., {Gandolfi}, D., {Armstrong}, D.~J., {et~al.} 2020, \mnras, 492, 5399

\bibitem[{{Otegi} {et~al.}(2020){Otegi}, {Bouchy}, \& {Helled}}]{otegi2020_MR}
{Otegi}, J.~F., {Bouchy}, F., \& {Helled}, R. 2020, \aap, 634, A43

\bibitem[{{Owen} \& {Wu}(2017)}]{owen_wu2017}
{Owen}, J.~E. \& {Wu}, Y. 2017, \apj, 847, 29

\bibitem[{Pascanu {et~al.}(2013)Pascanu, Gulcehre, Cho, \& Bengio}]{pascanu2013construct}
Pascanu, R., Gulcehre, C., Cho, K., \& Bengio, Y. 2013, arXiv:1312.6026

\bibitem[{{Paxton} {et~al.}(2011){Paxton}, {Bildsten}, {Dotter}, {Herwig}, {Lesaffre}, \& {Timmes}}]{paxton2011}
{Paxton}, B., {Bildsten}, L., {Dotter}, A., {et~al.} 2011, \apjs, 192, 3

\bibitem[{{Paxton} {et~al.}(2013){Paxton}, {Cantiello}, {Arras}, {Bildsten}, {Brown}, {Dotter}, {Mankovich}, {Montgomery}, {Stello}, {Timmes}, \& {Townsend}}]{paxton2013}
{Paxton}, B., {Cantiello}, M., {Arras}, P., {et~al.} 2013, \apjs, 208, 4

\bibitem[{{Paxton} {et~al.}(2015){Paxton}, {Marchant}, {Schwab}, {Bauer}, {Bildsten}, {Cantiello}, {Dessart}, {Farmer}, {Hu}, {Langer}, {Townsend}, {Townsley}, \& {Timmes}}]{paxton2015}
{Paxton}, B., {Marchant}, P., {Schwab}, J., {et~al.} 2015, \apjs, 220, 15

\bibitem[{{Paxton} {et~al.}(2018){Paxton}, {Schwab}, {Bauer}, {Bildsten}, {Blinnikov}, {Duffell}, {Farmer}, {Goldberg}, {Marchant}, {Sorokina}, {Thoul}, {Townsend}, \& {Timmes}}]{paxton2018}
{Paxton}, B., {Schwab}, J., {Bauer}, E.~B., {et~al.} 2018, \apjs, 234, 34

\bibitem[{Powell(2001)}]{powell2001radial}
Powell, M. 2001, in HERCMA, Citeseer, 2--24

\bibitem[{Ramachandran {et~al.}(2017)Ramachandran, Zoph, \& Le}]{ramachandran2017searching}
Ramachandran, P., Zoph, B., \& Le, Q.~V. 2017, arXiv:1710.05941

\bibitem[{{Ricker} {et~al.}(2015){Ricker}, {Winn}, {Vanderspek}, {Latham}, {Bakos}, {Bean}, {Berta-Thompson}, {Brown}, {Buchhave}, {Butler}, {Butler}, {Chaplin}, {Charbonneau}, {Christensen-Dalsgaard}, {Clampin}, {Deming}, {Doty}, {De Lee}, {Dressing}, {Dunham}, {Endl}, {Fressin}, {Ge}, {Henning}, {Holman}, {Howard}, {Ida}, {Jenkins}, {Jernigan}, {Johnson}, {Kaltenegger}, {Kawai}, {Kjeldsen}, {Laughlin}, {Levine}, {Lin}, {Lissauer}, {MacQueen}, {Marcy}, {McCullough}, {Morton}, {Narita}, {Paegert}, {Palle}, {Pepe}, {Pepper}, {Quirrenbach}, {Rinehart}, {Sasselov}, {Sato}, {Seager}, {Sozzetti}, {Stassun}, {Sullivan}, {Szentgyorgyi}, {Torres}, {Udry}, \& {Villasenor}}]{ricker2015_tess}
{Ricker}, G.~R., {Winn}, J.~N., {Vanderspek}, R., {et~al.} 2015, Journal of Astronomical Telescopes, Instruments, and Systems, 1, 014003

\bibitem[{{Rogers} \& {Seager}(2010)}]{rogers_seager2010}
{Rogers}, L.~A. \& {Seager}, S. 2010, \apj, 712, 974

\bibitem[{Salehinejad {et~al.}(2017)Salehinejad, Sankar, Barfett, Colak, \& Valaee}]{salehinejad2017recent}
Salehinejad, H., Sankar, S., Barfett, J., Colak, E., \& Valaee, S. 2017, arXiv:1801.01078

\bibitem[{{Sanz-Forcada} {et~al.}(2011){Sanz-Forcada}, {Micela}, {Ribas}, {Pollock}, {Eiroa}, {Velasco}, {Solano}, \& {Garc{\'\i}a-{\'A}lvarez}}]{SF2011}
{Sanz-Forcada}, J., {Micela}, G., {Ribas}, I., {et~al.} 2011, \aap, 532, A6

\bibitem[{Sharma {et~al.}(2017)Sharma, Sharma, \& Athaiya}]{sharma2017activation}
Sharma, S., Sharma, S., \& Athaiya, A. 2017, Towards Data Sci, 6, 310

\bibitem[{{Shkolnik} \& {Barman}(2014)}]{Shkolnik2014}
{Shkolnik}, E.~L. \& {Barman}, T.~S. 2014, \aj, 148, 64

\bibitem[{Sibi {et~al.}(2013)Sibi, Jones, \& Siddarth}]{sibi2013analysis}
Sibi, P., Jones, S.~A., \& Siddarth, P. 2013, Journal of theoretical and applied information technology, 47, 1264

\bibitem[{{Spada} {et~al.}(2013){Spada}, {Demarque}, {Kim}, \& {Sills}}]{spada2013}
{Spada}, F., {Demarque}, P., {Kim}, Y.~C., \& {Sills}, A. 2013, \apj, 776, 87

\bibitem[{{St{\"o}kl} {et~al.}(2015){St{\"o}kl}, {Dorfi}, \& {Lammer}}]{stoekl2015}
{St{\"o}kl}, A., {Dorfi}, E., \& {Lammer}, H. 2015, \aap, 576, A87

\bibitem[{{Tuson} {et~al.}(2023){Tuson}, {Queloz}, {Osborn}, {Wilson}, {Hooton}, {Beck}, {Lendl}, {Olofsson}, {Fortier}, {Bonfanti}, {Brandeker}, {Buchhave}, {Collier Cameron}, {Ciardi}, {Collins}, {Gandolfi}, {Garai}, {Giacalone}, {Gomes da Silva}, {Howell}, {Patel}, {Persson}, {Serrano}, {Sousa}, {Ulmer-Moll}, {Vanderburg}, {Ziegler}, {Alibert}, {Alonso}, {Anglada}, {B{\'a}rczy}, {Barrado Navascues}, {Barros}, {Baumjohann}, {Beck}, {Benz}, {Billot}, {Bonfils}, {Borsato}, {Broeg}, {Cabrera}, {Charnoz}, {Conti}, {Csizmadia}, {Cubillos}, {Davies}, {Deleuil}, {Delrez}, {Demangeon}, {Demory}, {Dragomir}, {Dressing}, {Ehrenreich}, {Erikson}, {Essack}, {Farinato}, {Fossati}, {Fridlund}, {Furlan}, {Gill}, {Gillon}, {Gnilka}, {Gonzales}, {G{\"u}del}, {G{\"u}nther}, {Hoyer}, {Isaak}, {Jenkins}, {Kiss}, {Laskar}, {Latham}, {Law}, {Lecavelier des Etangs}, {Curto}, {Lovis}, {Luque}, {Magrin}, {Mann}, {Maxted}, {Mayor}, {McDermott}, {Mecina}, {Mordasini}, {Mortier}, {Nascimbeni}, {Ottensamer}, {Pagano}, {Pall{\'e}},
  {Peter}, {Piotto}, {Pollacco}, {Pritchard}, {Ragazzoni}, {Rando}, {Ratti}, {Rauer}, {Ribas}, {Ricker}, {Rieder}, {Santos}, {Savel}, {Scandariato}, {Schwarz}, {Seager}, {S{\'e}gransan}, {Shporer}, {Simon}, {Smith}, {Steller}, {Stockdale}, {Szab{\'o}}, {Thomas}, {Torres}, {Tronsgaard}, {Udry}, {Ulmer}, {Van Grootel}, {Vanderspek}, {Venturini}, {Walton}, {Winn}, \& {Wohler}}]{tuson2023_hd15906}
{Tuson}, A., {Queloz}, D., {Osborn}, H.~P., {et~al.} 2023, \mnras, 523, 3090

\bibitem[{{Venturini} {et~al.}(2020){Venturini}, {Guilera}, {Haldemann}, {Ronco}, \& {Mordasini}}]{venturini2020}
{Venturini}, J., {Guilera}, O.~M., {Haldemann}, J., {Ronco}, M.~P., \& {Mordasini}, C. 2020, \aap, 643, L1

\bibitem[{{Watson} {et~al.}(1981){Watson}, {Donahue}, \& {Walker}}]{watson1981}
{Watson}, A.~J., {Donahue}, T.~M., \& {Walker}, J.~C.~G. 1981, \icarus, 48, 150

\bibitem[{{Wilson} {et~al.}(2022){Wilson}, {Goffo}, {Alibert}, {Gandolfi}, {Bonfanti}, {Persson}, {Collier Cameron}, {Fridlund}, {Fossati}, {Korth}, {Benz}, {Deline}, {Flor{\'e}n}, {Guterman}, {Adibekyan}, {Hooton}, {Hoyer}, {Leleu}, {Mustill}, {Salmon}, {Sousa}, {Suarez}, {Abe}, {Agabi}, {Alonso}, {Anglada}, {Asquier}, {B{\'a}rczy}, {Barrado Navascues}, {Barros}, {Baumjohann}, {Beck}, {Beck}, {Billot}, {Bonfils}, {Brandeker}, {Broeg}, {Bryant}, {Burleigh}, {Buttu}, {Cabrera}, {Charnoz}, {Ciardi}, {Cloutier}, {Cochran}, {Collins}, {Col{\'o}n}, {Crouzet}, {Csizmadia}, {Davies}, {Deleuil}, {Delrez}, {Demangeon}, {Demory}, {Dragomir}, {Dransfield}, {Ehrenreich}, {Erikson}, {Fortier}, {Gan}, {Gill}, {Gillon}, {Gnilka}, {Grieves}, {Grziwa}, {G{\"u}del}, {Guillot}, {Haldemann}, {Heng}, {Horne}, {Howell}, {Isaak}, {Jenkins}, {Jensen}, {Kiss}, {Lacedelli}, {Lam}, {Laskar}, {Latham}, {Lecavelier des Etangs}, {Lendl}, {Lester}, {Levine}, {Livingston}, {Lovis}, {Luque}, {Magrin}, {Marie-Sainte}, {Maxted}, {Mayo},
  {McLean}, {Mecina}, {M{\'e}karnia}, {Nascimbeni}, {Nielsen}, {Olofsson}, {Osborn}, {Osborne}, {Ottensamer}, {Pagano}, {Pall{\'e}}, {Peter}, {Piotto}, {Pollacco}, {Queloz}, {Ragazzoni}, {Rando}, {Rauer}, {Redfield}, {Ribas}, {Ricker}, {Rieder}, {Santos}, {Scandariato}, {Schmider}, {Schwarz}, {Scott}, {Seager}, {S{\'e}gransan}, {Serrano}, {Simon}, {Smith}, {Steller}, {Stockdale}, {Szab{\'o}}, {Thomas}, {Ting}, {Triaud}, {Udry}, {Van Eylen}, {Van Grootel}, {Vanderspek}, {Viotto}, {Walton}, \& {Winn}}]{wilson2022_toi1264}
{Wilson}, T.~G., {Goffo}, E., {Alibert}, Y., {et~al.} 2022, \mnras, 511, 1043

\bibitem[{Wu(2017)}]{wu2017introduction}
Wu, J. 2017, National Key Lab for Novel Software Technology. Nanjing University. China, 5, 495

\bibitem[{Wu \& Feng(2018)}]{wu2018development}
Wu, Y.-c. \& Feng, J.-w. 2018, Wireless Personal Communications, 102, 1645

\bibitem[{{Yelle} {et~al.}(2008){Yelle}, {Lammer}, \& {Ip}}]{yelle2008}
{Yelle}, R., {Lammer}, H., \& {Ip}, W.-H. 2008, \ssr, 139, 437

\bibitem[{{Yelle}(2004)}]{yelle2004Icar..170..167Y}
{Yelle}, R.~V. 2004, \icarus, 170, 167

\end{thebibliography}

\begin{appendix}

\section{Relation of hydrodynamic model predictions to common analytical approximation}
\label{apx:EL-CP-RLO}
In different regions of the parameter space, hydrodynamic escape can be driven by photoionisation heating (XUV-driven escape), the own thermal energy of a planet (core-powered mass loss or boil-off), tidal forces of the host star (Roche lobe overflow), or their combination. To help understanding which of the mechanisms is dominant for a specific planet, it is convenient to compare the predictions of hydrodynamic models to analytic approximations parametrising each of these processes separately.
\subsection{Energy-limited approximation}
\begin{figure} 
\centering
\includegraphics[width=1\linewidth]{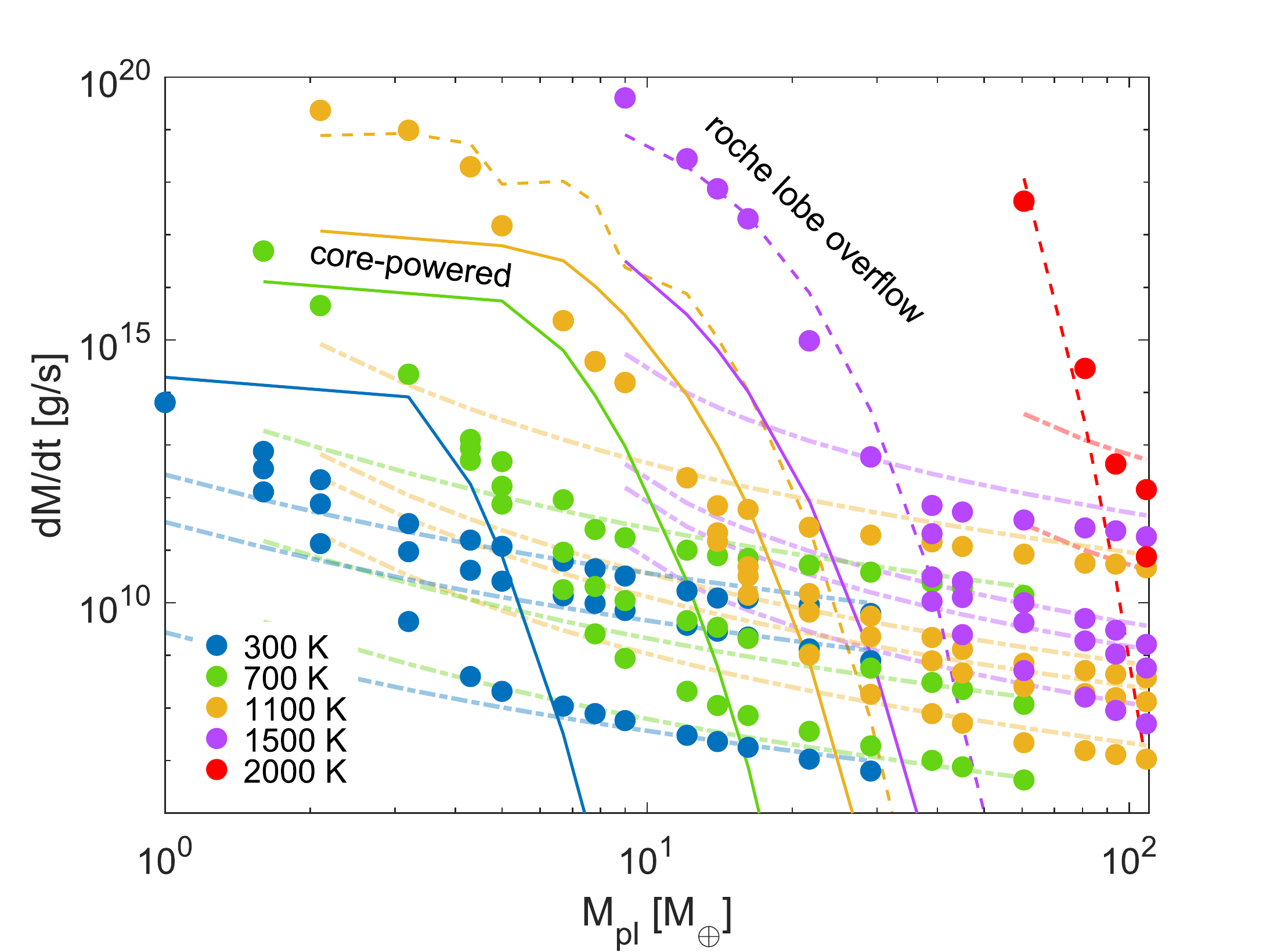}
\includegraphics[width=1\linewidth]{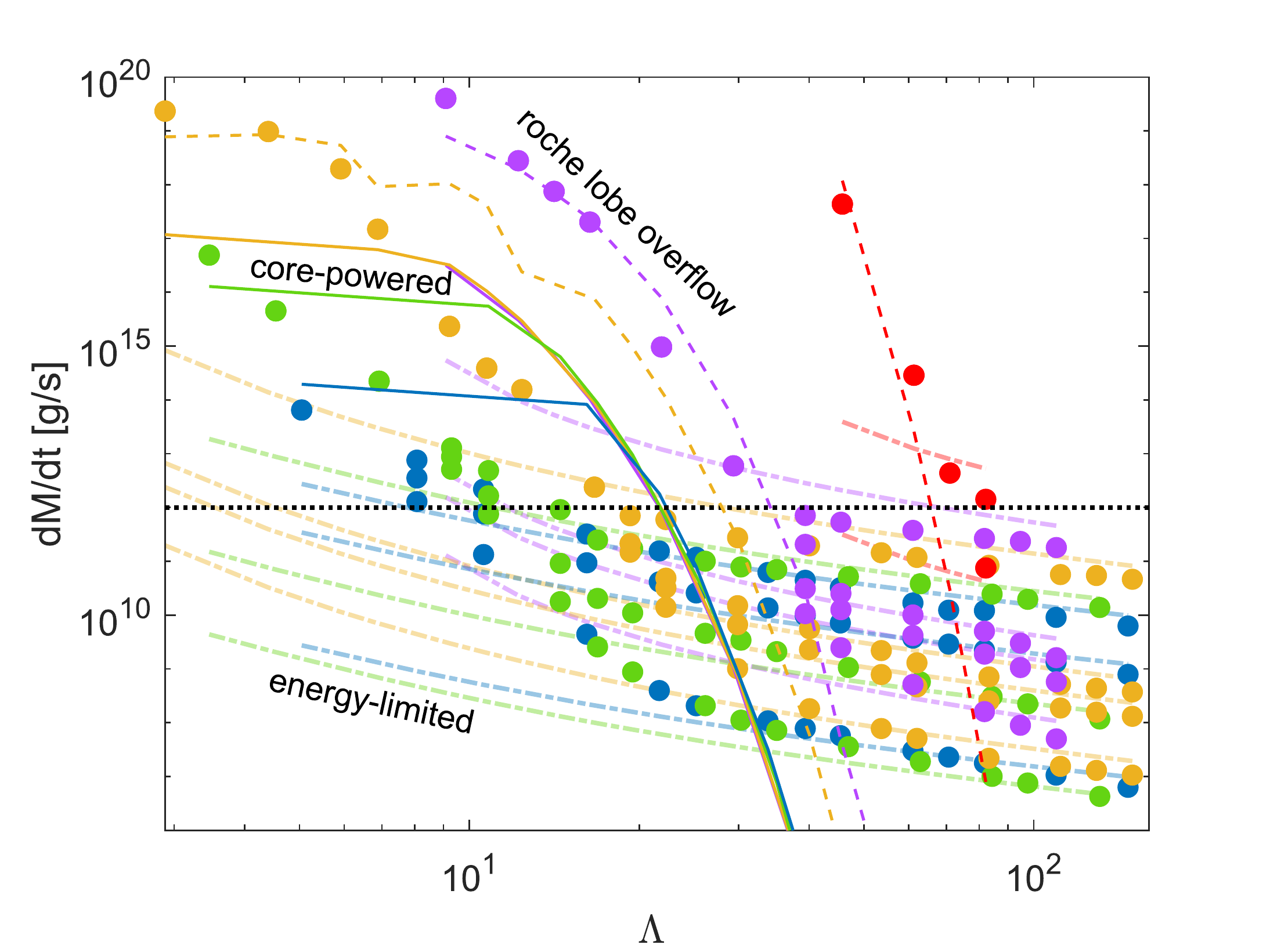}
\includegraphics[width=1\linewidth]{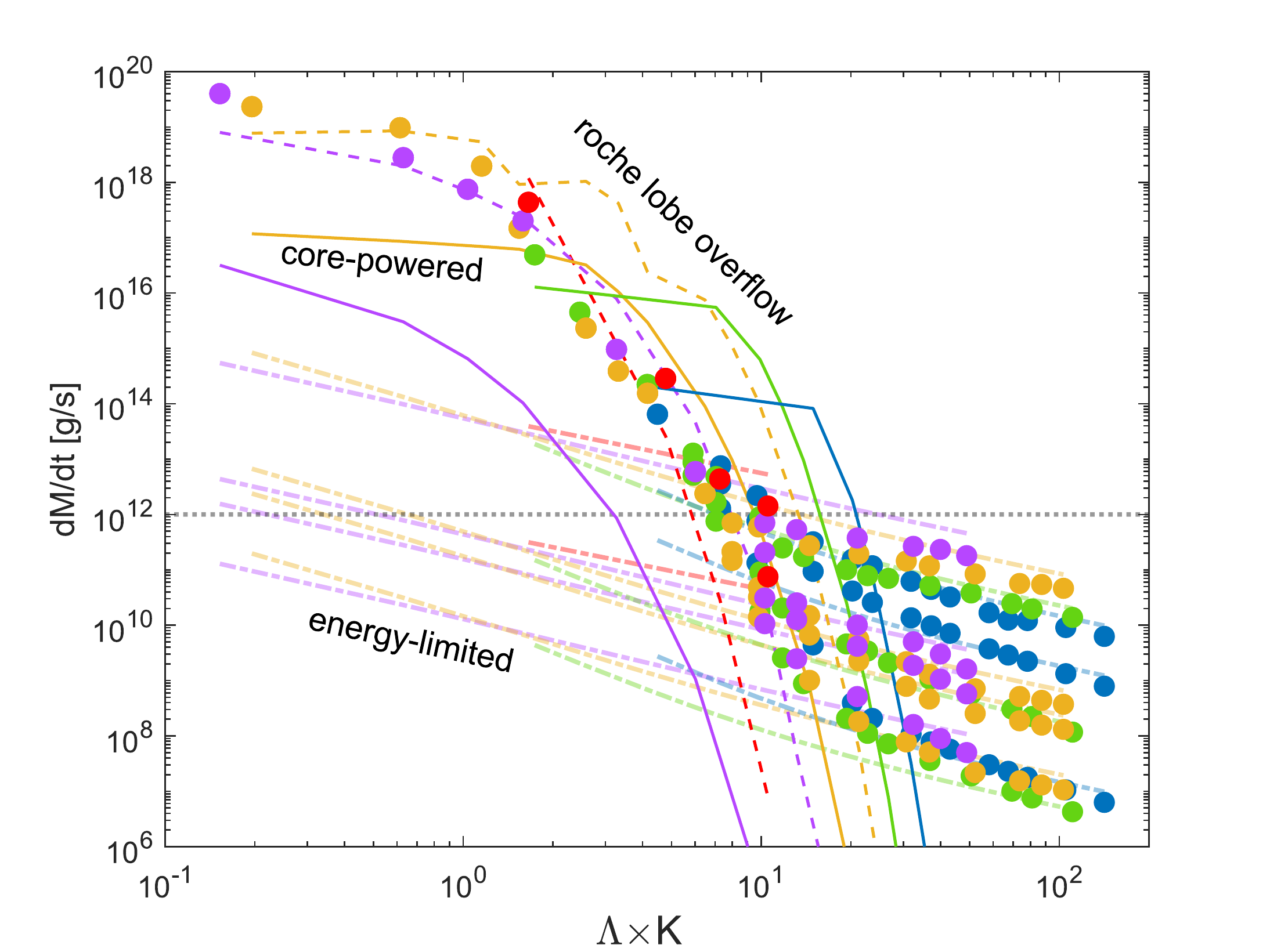}
\caption{Same as in Figure\,\ref{fig:LHY-PROCESSES_CLEAN}, but with more \Feuv\ values. Multiple dash-dotted lines of the same colour show the predictions of the energy-limited approximation for different \Feuv\ values, and the multiple sequences of circles of the same colours give the predictions of the hydrodynamic model for the corresponding \Feuv\ values (the latter coincide for the low-gravity planets). The considered EUV emission values are (from the bottom to the top line/sequence of each colour): 3.0, 375.5, and 3000\,\ergscm\ for 300\,K; 2.6, 89.1, and $\sim1.1\times10^4$\,\ergscm\ for 700\,K; 16, 193.6, 543.3, and $\sim6.8\times10^4$\,\ergscm\ for 1100\,K; 55.4, 669.6, 1878.6, and $\sim2.35\times10^5$\,\ergscm\ for 1500\,K; 175.2, 5937.3, and $\sim7.4\times10^5$\,\ergscm\ for 2000\,K.}
\label{fig:LHY-PROCESSES}
\end{figure}

For XUV-driven escape, we consider the energy-limited approximation (see dash-dotted lines in Figure\,\ref{fig:LHY-PROCESSES})
\begin{equation}\label{eq:ELim}
    \dot{M}_{\rm EL} = \eta\frac{\Rpl R_{\rm eff}^2\Feuv}{KG\Mpl}\,,
\end{equation}
where $\eta$ is the heating efficiency quantifying the fraction of the total stellar energy absorbed in the atmosphere transformed into heating, $G$ is the gravitational constant, and $K$ is the parameter quantifying the effect of stellar tidal forces \citep{Erkaev2007} (see Equation\,\ref{eq:K}).
The effective radius of EUV absorption (located somewhat above the photosphere) was taken according to the approximation \citep{chen_rogers2016}
\begin{equation}
\label{eq:Reff}
R_{\rm eff} = 1 + H\log\left(\frac{P_{\rm photo}}{P_{\rm EUV}}\right)\,,
\end{equation}
where $H$ is the atmospheric height scale normalised to the planetary radius
\begin{equation}
H = \frac{k_{\rm b}\Teq}{2m_{\rm H}g\Rpl}\,,
\end{equation}
$P_{\rm EUV}$ estimates the pressure at the EUV absorption level
\begin{equation}
P_{\rm EUV} = \frac{m_{\rm H}g}{\sigma_{\rm 20\,eV}}\,,\,\,\,
\sigma_{\rm 20\,eV} = 6.0\times10^{-18}\left(\frac{20\,{\rm eV}}{13.6\,{\rm eV}}\right)\,,
\end{equation}
$g\,=\,G\Mpl/\Rpl^2$ is the planetary gravitational potential, $k_{\rm b}$ is the Boltzmann constant, and $\sigma_{\rm 20\,eV}$ is the absorption cross-section of photons with energy of 20\,eV. The photospheric pressure $P_{\rm photo}$ was set to the values employed for the lower boundary of the simulation domain in the hydrodynamic models (Equation\,\ref{eq:PRESSURE0}).

\subsection{Core-powered mass loss}
To assess how the predictions of the hydrodynamic models compare with the core-powered mass loss approximation, we follow the approach of \citet{gupta_schlichting2019} and make the following assumptions. First, we assume that the atmospheres have solar metallicity ($Z_{*}=1$) and behave as an ideal gas (adiabatic index $\gamma=5/3$). We also assume that the mean weight of the atmospheric species is approximately the weight of molecular hydrogen ($2m_{\rm H}$) at the planetary photosphere and the weight of atomic hydrogen ($m_{\rm H}$) at high altitudes (e.g. at the sonic point/Bondi radius). 
For the radius of the planetary core (i.e. solid part of the planet composed of rocks and metals), we adopt the approximation \citep{rogers_seager2010}
\begin{equation}\label{eq:Rcore}
\frac{R_{\rm core}}{\Rer} = \left(\frac{\Mpl}{\Mer}\right)^{0.27}
\end{equation}
and assume that the core mass $M_{\rm core}$ is approximately equal to \Mpl. Both assumptions on the core parameters have a minor impact on the results. Differently from the hydrodynamic models, here gravity is defined at the core radius, namely $g_{\rm core} = GM_{\rm core}/R_{\rm core}^2$.

Furthermore, \citet{gupta_schlichting2019} assume that the radiative-convective boundary (RCB) coincides approximately with the visible radius ($R_{\rm RCB}\,\simeq\,\Rpl$, which might not be strictly accurate for puffy sub-Neptunes). In this case, the temperature at the RCB can be assumed being $T_{\rm RCB}\,\simeq\,\Teq$ (assuming zero albedo, as for the grid planets), and for the pressure at the RCB ($P_{\rm RCB}$) one can employ the values predicted by Equation~\ref{eq:PRESSURE0}. Therefore, the parameters adopted for the RCB are identical to the lower boundary parameters of the hydrodynamic models. Then, the density at RCB can be computed as
\begin{equation}
\rho_{\rm RCB} = \rho_{\rm ph} = \frac{m_{\rm H}P_{\rm rcb}}{k_{\rm b}\,T_{\rm rcb}}\,.
\end{equation}
The sonic velocity at the photosphere is
$C_{\rm s0} = \sqrt{k_{\rm b}\Teq/m_{\rm H}}$, and thus under the assumption of isothermal atmosphere the sonic point can be estimated as 
\begin{equation}
R_{\rm s} = \frac{G\Mpl}{2C_{\rm s0}^2}\,.
\end{equation}
The modified Bondi radius \citep{ginzburg2018} is
\begin{equation}
R_{\rm B} = \frac{\gamma-1}{\gamma}\frac{Gm_{\rm H}*M_{\rm core}}{k_{\rm b}T_{rcb}}\,,
\end{equation}
and the opacity at the radiative-convective boundary is given by the approximation \citep{freedman2008,lee_chiang2015ApJ...811...41L}
\begin{equation}
\kappa = 0.1Z_*\left(\frac{\rho_{\rm rcb}}{1e-3}\right).^{0.6}\,.
\end{equation}
Following these assumptions and considerations, the cooling luminosity of a planet can be estimated as
\begin{equation}
L_{\rm pl} = \frac{64\pi}{3}\frac{\sigma_{\rm SB}T_{\rm RCB}^4R_{\rm B}}{\kappa\rho_{\rm RCB}}\,,
\end{equation}
where $\sigma_{\rm SB}\,=\,5.6704\times10^{-5}\,{\rm erg\,cm^{-2}\,s^{-1}\,K^{-4}}$ is the Stefan-Boltzmann constant. Assuming the entire cooling luminosity goes into driving the escape, the mass loss powered by the core luminosity is
\begin{equation}
\dot{M}_{\rm CL} = \frac{L_{\rm pl}}{gR_{\rm core}}
\end{equation}
This value is further restricted by the mechanical limit on the atmospheric mass-loss rate set by the thermal velocity of the gas molecules at the Bondi radius, which is considered to be approximately equal to that at the sonic point \citep[][Bondi-limited mass-loss]{Ginzburg2016}
\begin{equation}
\dot{M}_{\rm B} = 4\pi R_{\rm s}^2C_{\rm s0}\rho_{\rm RCB}\exp(-\frac{GM_{\rm core}}{\Rpl C_{\rm s0}^2})\,.
\end{equation}
The actual core-powered mass-loss rate $\dot{M}_{\rm CP}$ is then given by the minimum of the two mass-loss rates $\dot{M}_{\rm CL}$ and $\dot{M}_{\rm BL}$.

\subsection{Roche lobe overflow limit}
    
To roughly estimate mass-loss rates due to atmospheric disruption by stellar tidal forces, we modify the analytical expression of \citet{Koskinen2022ApJ...929...52K}. The original formulation quantifies the mass-loss rate through scaling based on the difference in the gravitational potentials between the Roche lobe and the exobase. However, in general, one needs to employ hydrodynamic modelling to know the position and other parameters of the exobase. To overcome this uncertainty, we use the difference between the Roche lobe and the photosphere, which is also indicative of the effect of stellar gravity on a planetary atmosphere \citep[as was used to derive $K$;][]{Erkaev2007}. In this way, atmospheric mass loss due to Roche lobe overflow can be approximated as 
\begin{equation}\label{eq:Mdot_RLO}
\dot{M}_{\rm RL} = 2\sqrt{2\pi}\Rpl\hat{R}_{\rm pl}\rho_{\rm ph}C_{\rm s0}\left(1 + G_{\rm roc}\right)\exp(-G_{\rm roc})\,,
\end{equation}
where the first part of the expression on the right hand side quantifies the mass flow through the photosphere, if it would occur with sonic velocity $C_{\rm s0}$ (which is close to reality in extreme cases), and $G_{\rm roc}$ is proportional to the difference in gravitational potential $U_{\rm r}$ between the photosphere and the Roche lobe 
\begin{equation}
G_{\rm roc} = \frac{(U_{\rm r,exo}-U_{\rm r,roc})}{C_{\rm s0}^2}\,,
\end{equation}
with the gravitational potential accounting for the gravitational forces of the planet and of the host star, and for the orbital motion
\begin{equation}\label{eq:gravpot_r}
U_{\rm r}(r) = \frac{G\Mpl}{r} - \frac{GM_*}{a-r} -  \frac{G(M_* + \Mpl)}{2a^3}\left(\frac{aM_*}{M_*+\Mpl}-r\right)^2\,.
\end{equation}
Finally, $\hat{R}_{\rm pl}$ in Equation\,\ref{eq:Mdot_RLO} accounts for the deviation of the equipotential surface from spherical form: it is the distance in polar direction $z$ where the gravitational potential $U_{\rm z}$ becomes equal to $U_{\rm r}(\Rpl)$, where
\begin{equation}
U_{\rm z}(z) = \frac{G\Mpl}{z} - \frac{G0\,Mstar}{\sqrt{a^2 + z^2}} - \frac{G(M_* + \Mpl)}{2a^3}\,\left(\frac{aM_*}{M_*+\Mpl}\right)^2\,.
\end{equation}   

Equation~\ref{eq:Mdot_RLO} overestimates the escape for the majority of planets, as it assumes that the atmospheric outflow occurs with a speed close to the sonic velocity already at the photosphere, while for a typical planet in the XUV-driven regime within our grid the outflow velocity at the photosphere is close, or equal, to zero. However, it can reproduce some of the most extreme points in the grid (hot puffy planets with $\Rroc\, \lesssim \, 1.8 \, \Rpl$. 
\section{Quantifying the quality of data coverage}\label{apx:troublemakers}
We first define the parameter differences between test points and the nearest neighbouring grid points, hierarchically, starting from $M_*$, and proceeding with \Teq, \Feuv, and \Rpl. For a test point with parameters $x^{\rm i}\,=\,\{M_*^{\rm i}, \Teq^{\rm i}, a^{\rm i}, \Feuv^{\rm i}, \Rpl^{\rm i}, \Mpl^{\rm i}\}$, we first define the two stellar masses in the grid where $|M_*^{\rm g}\,-\,M_*^{\rm i}|$ is the smallest. Typically, $M_*^{\rm i}$ is located between the two grid values $M_*^{\rm g,1}$ and $M_*^{\rm g,2}$. We then consider all the \Teq\ values available in the grid for $M_*^{\rm g,1}$ and $M_*^{\rm g,2}$ and introduce the parameter $J_{\rm bord}^{Teq,j}$, with $j\,=\,1,2$, setting it equal to 0, if $\Teq^{\rm i}$ is between the maximum and the minimum temperature available for $M_*^{\rm g,j}$, and setting it equal to 1 otherwise. Then, for each $M_*^{\rm g,1}$ and $M_*^{\rm g,2}$, we define the two equilibrium temperatures closest to $\Teq^{\rm i}$, obtaining four differences between the test point $i$ and the nearest neighbouring grid points in terms of \Teq\ ($\Teq^{\rm g,11}$, $\Teq^{\rm g,12}$, $\Teq^{\rm g,21}$, $\Teq^{\rm g,22}$). The differences in terms of $a$ are the same, as \Teq\ and $a$ are directly linked. For each $\Teq^{\rm g,jk}$, with $k\,=\,1,2$, we consider the range of available \Feuv\ values and define $J_{\rm bord}^{EUV,jk}\,=\,(1|0)$ similarly to $J_{\rm bord}^{Teq,j}$.
Similarly, we consider the \Feuv\ values closest to $\Feuv^{\rm i}$ for each pair of $M_*^{\rm g}$ and $\Teq^{\rm g}$, obtaining eight differences between the test point $i$ and the nearest neighbouring grid points, and proceed in the same way for \Rpl\ for each pair of $M_*^{\rm g}$, $\Teq^{\rm g}$, and $\Feuv^{\rm g}$, obtaining 16 differences between the test point $i$ and the nearest neighbouring grid points, and thus defining $J_{\rm bord}^{Rpl,jkl}$ and $J_{\rm bord}^{Mpl,jklm}$ ($l,m\,=\,1,2$). Finally, similarly to $J_{\rm bord}^{Teq,j}$ we define its analogue in terms of orbital separation $J_{\rm bord}^{a,j}$, and similarly to $J_{\rm bord}^{Mpl,jklm}$ we define the $J_{\rm bord}$ parameters based on $\Lambda$ and $\Lambda\times K$ ($J_{\rm bord}^{\Lambda,jklm}$ and $J_{\rm bord}^{\Lambda K,jklm}$). At last, we compute the cumulative $J_{\rm bord}$ value as
\begin{eqnarray}
J_{\rm bord} &=& \frac{1}{8}\left( J_{\rm bord}^{M_*} + \frac{1}{2}\sum_{\rm j=1}^2{J_{\rm bord}^{Teq,j}} + \frac{1}{2}\sum_{\rm j=1}^2{J_{\rm bord}^{a,j}} \nonumber\right.\\ 
&+&  \frac{1}{4}\sum_{\rm j,k=1}^2{J_{\rm bord}^{EUV,jk}} + \frac{1}{8}\sum_{\rm j,k,l=1}^2{J_{\rm bord}^{Rpl,jkl}} + \frac{1}{16}\sum_{\rm j,k,l,m=1}^2{J_{\rm bord}^{Mpl,jklm}} \nonumber\\ 
&+&  \left.\frac{1}{16}\sum_{\rm j,k,l,m=1}^2{J_{\rm bord}^{\Lambda,jklm}} + \frac{1}{16}\sum_{\rm j,k,l,m=1}^2{J_{\rm bord}^{\Lambda K,jklm}} \right) \,.
\end{eqnarray}
\section{Optimisation on RBF kernels} 
\label{apx:rbf-kernel}
We comprehensively explored the available kernel options in the ``RBFInterpolator'' function of the $\texttt{SciPy}$ package. These kernels include $\textit{linear}$, $\textit{gaussian}$, $\textit{cubic}$, $\textit{thin_plate_spline}$, $\textit{quintic}$, $\textit{multiquadric}$, $\textit{inverse_multiquadric}$ ($\text{inv_mul}$), and $\textit{inverse_quadratic}$. We used a grid-based search approach to optimize the shape parameter ($\boldsymbol{\epsilon}$) and the degree of the polynomial ($\boldsymbol{d}$) where applicable. We defined a grid of values for each kernel to test the shape parameter and polynomial degrees within a suitable range. The polynomial degrees for applicable kernels have been varied between $1$ and $4$.

The ranges of shape parameters varied depending on the kernel due to their differing mathematical characteristics. For instance: (a) For the quintic kernel, we tested $\boldsymbol{\epsilon}$ values from the grid $[0.01, 0.02, 0.1, 0.2, 0.3, 1.0, 2.0, 3.0, 5.0, 10.0]$. (b) The shape parameter does not considerably influence the smoothing behaviour for the linear kernel. However, for a thorough investigation, we evaluated a range of values $[0.1, 0.2, 0.3, 0.4, 0.5, 0.6, 1.0, 2.0, 3.0, 5.0, 10.0]$. After a grid search, we found that the best shape parameter value was $1.0$, aligning with theoretical expectations. Table~\ref{tab:RBF_params} shows the optimised values of $\boldsymbol{\epsilon}$ and $\boldsymbol{d}$ for each kernel.

\begin{table}
\centering
\caption{Optimised $\boldsymbol{\epsilon}$ and $\boldsymbol{d}$ values after grid-based search.}
\begin{tabular}{lcc}
\hline
\hline
Kernel type & $\boldsymbol{\epsilon}$ & $\boldsymbol{d}$ \\
\hline
$\textit{linear}$              & 1     & 1 \\
$\textit{cubic}$               & 0.4   & 1 \\
$\textit{multiquadric}$        & 10    & 1 \\
$\textit{inverse_multiquadric}$ & 3     & 1 \\
$\textit{thin_plate_spline}$   & 3     & 1 \\
$\textit{inverse_quadratic}$   & 2     & 1 \\
$\textit{gaussian}$            & 1     & 1 \\
$\textit{quintic}$             & 0.002 & 2 \\
\hline
\end{tabular}
\label{tab:RBF_params}
\end{table}

Except for $\textit{quintic}$ and $\textit{gaussian}$, all kernels ($\textit{linear}$, $\textit{cubic}$, $\textit{thin_plate_spline}$, $\textit{multiquadric}$, $\textit{inverse_multiquadric}$, and $\textit{inverse_quadratic}$) exhibited nearly similar accuracy with the test data. Despite the optimised kernel choice, none of these kernels outperformed the accuracy of the three-layer NN, indicating that DNN achieves slightly better accuracy than other RBF kernels for the test dataset.
\section{LSTM-network} 
\label{apx:lstm}
\begin{figure}[!ht]
\centering
\includegraphics[trim={0 0 0 0}, clip, scale = 0.66]{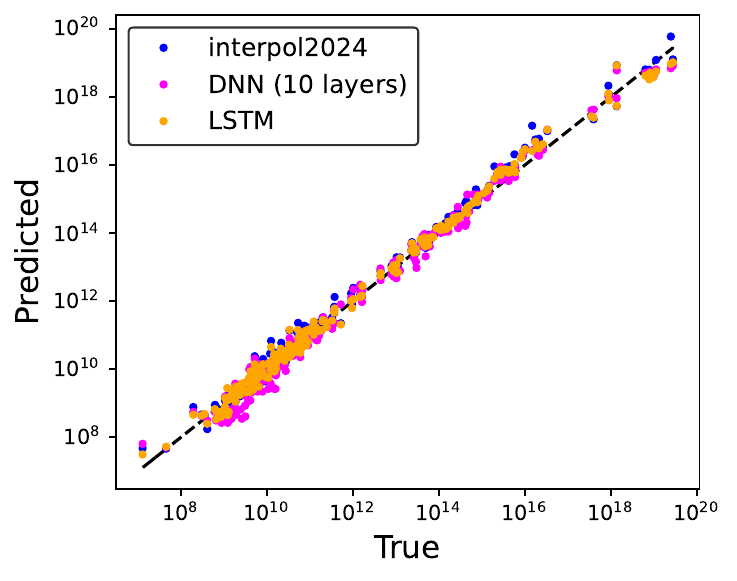} \hfill
\vspace{-0.3cm}
\caption{Mass-loss rates predicted using the LSTM (dark yellow), DNN (magenta), and \texttt{interpol2024} (blue) schemes, compared to the actual values obtained from the hydrodynamic modelling for the test data set. The LSTM architecture consists of a single LSTM layer with 50 units, while the DNN comprises ten hidden layers with 512 neurons each. The DNN architecture is significantly more complex than the LSTM, resulting in approximately 2.63 million trainable parameters for the DNN compared to around 11,500 for the LSTM. This heavier model impacts the DNN's overall accuracy, making it less accurate than the LSTM due to overfitting \citep{hinton2012improving}. However, with an extended dataset, it is expected that the DNN would perform better on unseen test data.}
\label{fig:pred_lstm}
\end{figure}

The DNN model captures the mass-loss rate calculation using six input parameters. Also, we explored Long Short-Term Memory (LSTM) networks for the same task. LSTM, a specialised recurrent neural network (RNN), uses cell states, hidden states, and gating mechanisms to process sequential data and produce continuous outputs \citep{hochreiter1997long}. The model includes one LSTM layer with 50 units, designed to handle input sequences of shape (1,6), corresponding to single-time steps with six features, followed by a dense output layer generating a continuous output. Figure~\ref{fig:pred_lstm} shows the model's predictions. Multiple LSTM layers can be stacked to enhance performance, improving accuracy with more extensive grid data. We will explore deeper DNN architectures with additional layers (e.g. 10) and neurons (e.g. 512) or more complex Dense LSTM models as more comprehensive grids become available. We remark that a ten-layer DNN or LSTM architecture has the potential to achieve superior accuracy when trained with more extensive grid data, making them promising starting points for developing future advanced models. Meanwhile, the three-layer DNN and RBF models outperform the previously proposed method for the current training and testing dataset.
\end{appendix}


\end{document}